 \documentclass[onecolumn,12pt,draft]{IEEEtran} %!PN

\usepackage{epsf}

\begin{document}

%%\title{Using the Style File IEEEtran.sty}
%\title{
%Ultimate ``SIR'' (``Signal''-to-``Interference''-Ratio) in 
%Autonomous Linear Networks with Symmetric Weight Matrices, and Its Use to 
%Stabilize the Network 
%as Applied to Binary Associative Memory Systems
%} %!PN 

%\title{Ultimate ``SIR'' in Autonomous Linear Networks with Symmetric Weight Matrices, and Its Use to
%Stabilize the Network 
%%as Applied to Binary Associative Memory Systems
%} %!PN

\title{Ultimate ``SIR'' in Autonomous Linear Networks with Symmetric Weight Matrices, and Its Use to
Stabilize the Network - A Hopfield-like network
%as Applied to Binary Associative Memory Systems
} %!PN

\author{Zekeriya Uykan \thanks{Z. Uykan is with Do\u{g}u\c{s} University, Electronics and Communications 
Engineering, Acibadem, 34722 Kadik\"{o}y, Istanbul, Turkey. 
He was with the 
Helsinki University of Technology, Control 
Engineering Laboratory, FI-02015 HUT, Finland, and with Nokia Siemens Networks, Espoo, Finland. 
E-mail: zuykan@dogus.edu.tr. 
He was a visiting scientist at Harvard University Broadband Comm Lab., Cambridge, 02138, MA, 
from September 2008 to June 2009 
%to August 2009, 
and the first version of this work was performed during his stay at Harvard University.  
} 
}

%\markboth{IEEE Transactions On Neural Networks, Vol. XX, No. Y, Month 1999}
%{Murray and Balemi: Using the style file IEEEtran.sty} %!PN
%{Murray and Balemi: Using the Document Class IEEEtran.cls} %!PN

\maketitle
\begin{abstract}

%It's well-known that in a traditional continuous-time autonomous linear system, 
%the eigenvalues of the weight matrix solely determine the stability of the system. 
%If any of the eigenvalues is positive, then the system is unstable. In this 
%paper, we examine the linear system 
%with symmetric weight matrix which has at least one positive eigenvalue. 

In this paper, we present and analyse two Hopfield-like nonlinear networks, 
%called SALU-SIR and DSALU-SIR 
in continuous-time and discrete-time respectively. 
The proposed network is based on an autonomous linear system with a symmetric weight matrix, 
which is designed to be unstable, 
and a nonlinear function stabilizing the whole network 
thanks to a manipulated state variable called``ultimate SIR''.
This variable is observed to be equal to the traditional 
Signal-to-Interference Ratio (SIR) definition in telecommunications engineering. 
%in both continuousand discrete-time. 

The underlying linear system of the proposed continuous-time network is 
$\dot{ {\mathbf x}} = {\mathbf B} {\mathbf x}$ where 
{\bf B} is a real symmetric matrix whose 
diagonal elements are fixed to a constant. 
The nonlinear function, on the other hand, is based on the defined system variables 
called ``SIR''s. We also show that 
the ``SIR''s of all the states converge to a constant value, called ``system-specific Ultimate SIR''; 
which is equal to $\frac{r}{\lambda_{max}}$ where $r$ is the diagonal element
of matrix ${\bf B}$ 
and $\lambda_{max}$ is the maximum (positive) eigenvalue of diagonally-zero
matrix $({\bf B} - r{\bf I})$, where ${\bf I}$ denotes the identity matrix. 
The same result is obtained in its discrete-time version as well. 

Computer simulations for binary associative memory design problem show the effectiveness of the
proposed network as compared to the traditional Hopfield Networks.

\end{abstract}

\begin{keywords}
Autonomous Continuos/Discrete-Time Linear Systems, Hopfield Networks, Signal to Interference Ratio (SIR).
\end{keywords}

\section{Introduction \label{Section:INTRO}}

\PARstart{H}{opfield} Neural Networks \cite{Hopfield85} 
has been an important focus of 
research area in not only neural networks field but also circit and system theory 
since early 1980s whose applications vary from combinatorial 
optimization (e.g. \cite{Matsuda98}, \cite{Smith98} among many others) 
including travelling salesman problem (e.g. \cite{Tan05}, \cite{Huajin04} among others) 
to image restoration (e.g. \cite{Paik92}), 
from various control engineering optimization problems including robotics 
(e.g. \cite{Lendaris99} among others) to associative memory systems 
(e.g. \cite{Farrel90}, \cite{Muezzinoglu04} among others), etc. 
For a tutorial and further references about Hopfield NN, see e.g. \cite{Zurada92} and 
\cite{Haykin99}.

In this paper, we present and analyse two Hopfield-like nonlinear networks, called
SALU-SIR and DSALU-SIR in continuous-time and discrete-time respectively.

The proposed networks consist of linear and nonlinear parts: 
An autonomous linear system with a symmetric weight matrix, 
which is designed to be unstable, 
and a nonlinear function stabilizing the whole network.  
The underlying linear system of the proposed continuous time-network is 

%A traditional continuous-time autonomous dynamic linear network is given by  

\begin{equation} \label{eq:Diff_Linear_trad} 
\dot{ {\mathbf x}} =  {\mathbf B} {\mathbf x} 
\end{equation} 

where $\dot{ {\mathbf x}}$ shows the derivative of ${\mathbf x}$ with respect to time, i.e., 
$\dot{ {\mathbf x}} = \frac{d{\mathbf x}}{dt}$, and ${\mathbf B}$ is called system matrix or weight matrix.

We define the following system variables 
%called ``SIR'', 
denoted as $\theta_i$, 
%by rewriting the eq.(\ref{eq:cir}) in \cite{Uykan08a}: 

\begin{equation} \label{eq:cirA}
{\theta_i}(t) = \frac{ b_{ii} x_i(t)}{ d_i + \sum_{j = 1, j \neq i}^{N} b_{ij} x_j(t) },  
	\quad i=1, \dots, N
\end{equation}

where  
$x_i(t)$ represents the $i$'th neuron and  
$b_{ij}$ is the $ij$'th element of matrix ${\mathbf B}$. 
In this paper, we assume $d_i=0$ for the sake of brevity. 

We observe that the manipulated state variable ${\theta_i}(t)$ of eq.(\ref{eq:cirA}) resembles the 
well-known Signal-to-Interference-Ratio (SIR) definition in telecommunicaions engineering, 
which can be found in any related texbooks: 
Signal-to-Interference Ratio (SIR) is an important entity in commucations engineering 
which indicates the quality of a link between a transmitter and a receiver in a multi transmitter-receiver 
environment (see e.g. \cite{Rappaport96}, \cite{Pan08}).  
For example, let $N$ represent the number of transmitters and receivers using the same channel. Then the received 
SIR is given by (e.g. \cite{Rappaport96}) 

\begin{equation} \label{eq:cir} 
\gamma_i(t) = \frac{ g_{ii} p_i(t)}{ \nu_i + \sum_{j = 1, j \neq i}^{N} g_{ij} p_j(t) }, 
\quad i=1, \dots, N 
\end{equation} 

where  $p_i$ is the transmission power of transmitter $i$, $g_{ij}$ is
the link gain from transmitter $j$ to receiver $i$ (e.g. in case of wireless communications, 
$g_{ij}$ involves path loss, shadowing, etc) and $\nu_i$ is the
receiver noise at receiver $i$. 
Typically in communication systems like cellular radio systems, every transmitter tries to optimize its 
power $p_i$ such that the received SIR (i.e., $\gamma_i$) in eq.(\ref{eq:cir}) is kept at a 
target SIR value, $\gamma_i^{tgt}$. 

%In other words, the variable $\theta_i(k)$ for the system in 
%(\ref{eq:Diff_Linear_trad}) is inspired by the traditional SIR definition in (\ref{eq:cir}). 

It is well-known that the eigenvalues of the system matrix solely determine the stability of a linear 
dynamic networks. If any of the eigenvalues is positive, then the system is unstable. 
%Indeed, we design the linear part of the network to have positive eigenvalue(s). 
Designing the linear part of the network to have positive eigenvalue(s), 
we show,
%in section \ref{Section:USIR} 
in this paper, 
that the defined ``SIR'' in eq.(\ref{eq:cirA}) for any state 
appraches asymptotically to a constant, called ``ultimate SIR'', which is 
a function of the diagonal entry of the system matrix and its maximum eigenvalue. 
The nonlinear part of the network uses this result to stabilize the system. 
%in Section \ref{Section:proposedNet}. 
Finally, the proposed network is shown to exhibit features which are 
generally attributed to Hopfield Networks. 
Taking the sign of the converged states, the proposed network is applied to binary associative memory 
systems design.

%In this paper, we analyse the system with a 
%symmetric weight matrix whose diagonal elements are fixed to a constant and 
%which has at least one positive eigenvalue. 
%Using the same "SIR" definition in eq.(\ref{eq:cirA}) , 
%we, in this paper, show that the "SIR" of any state converges to a constant value, called "Ultimate SIR", in 
%such a network, and the "Ultimate SIR" is equal to $\frac{r}{\lambda_{max}}$ where $r$ is the diagonal element 
%of the weight matrix and $\lambda_{max}$ is the maximum (positive) eigenvalue of diagonally-zero 
%matrix $({\bf M} - r{\bf I})$, where ${\bf I}$ denotes the identity matrix. 

%In the later part of the paper, we use the introduced "Ultimate SIR" to 
%stabilize the (originally unstable) network. It's shown that the proposed 
%"Stabilized"-Autonomous-Linear-Networks-by-Ultimate-SIR" (SALU-SIR) 
%exhibits features which are 
%generally attributed to Hopfield Networks. 
%Taking the sign of the converged states, the SALU-SIR is applied to binary associative memory design. 

The paper is organized as follows:  The ``ultimate SIR" is analysed for the underlying linear dynamic part of the 
network with a symmetric weight matrix in section 
\ref{Section:USIR}.  Section \ref{Section:proposedNet} presents the stabilized network by the Ultimate "SIR" 
to be used as a binary associative memory system. 
Simulation results are presented in section \ref{Section:SimuResults} followed by the conclusions in 
Section \ref{Section:CONCLUSIONS}.

%\smallskip
\vspace{0.2cm}

%------------------------------------------------------------------------- 
%\section{Ultimate ``SIR'' in Autonomous Linear Networks with Symmetric Weight Matrix}
\section{Ultimate ``SIR'' Analysis \label{Section:USIR} }

In this section, we analyse the underlying linear system of the proposed network in both 
continuous-time and discrete-time respectively. 

%\vspace{2cm}

\subsection{ Continuous-time Analysis \label{Section:analysisCont} }

In this paper, we examine the case where weight matrix ${\mathbf B}$ in eq.(\ref{eq:Diff_Linear_trad}) 
is a real symmetric matrix whose diagonal elements are equal to a constant. So, 
the linear system in (\ref{eq:Diff_Linear_trad}) can be written as follows 

\begin{equation} \label{eq:Diff_Linear}
\dot{ {\mathbf x}} =  \Big( -r {\bf I} + {\mathbf W} \Big) {\bf x}
\end{equation}

where 
%$\dot{ {\mathbf x}}$ shows the derivative of ${\mathbf x}$ with respect to time, i.e., 
%$\dot{ {\mathbf x}} = \frac{d{\mathbf x}}{dt}$, and 

\begin{eqnarray} 
r {\bf I} =
\left[
\begin{array}{c c c c}
r   &   0   & \ldots  &  0 \\
0     &   r & \ldots  &  0 \\
\vdots &      & \ddots  &  0 \\
0     &   0   & \ldots  &  r
\end{array}
\right], 
\quad \quad  \textrm{and} \nonumber \\
{\mathbf W} =
\left[
\begin{array}{c c c c}
0  &   w_{12}   & \ldots  &  w_{1N} \\
w_{21}     &   0 & \ldots  &  w_{2N} \\
\vdots &     & \ddots  &  \vdots \\
w_{N1}    &   w_{N2}   & \ldots  &  0
\end{array}
\right]
%\quad \quad
%{\mathbf b} =
%\left[
%\begin{array}{c}
%b_1 \\
%b_2 \\
%\vdots \\
%b_N
%\end{array}
%\right]
\label{eq:matA_W_b}
\end{eqnarray}

where $w_{ij}=w_{ji}, \quad i,j=1,2, \dots, N$.

%The reasons of the notation in (\ref{eq:matA_W_b}) are twofold: 1) We prefer to have 
%the same notation as in \cite{Uykan08a}. 
%2) Later in the paper, we present a "stabilized" network which show similar features as Hopfield Network does, 
%and thus, the matrices $r{\mathbf I}$ and ${\mathbf W}$ are defined with the neural network terminology as follows: 
%Matrix $r{\mathbf I}$ shows the self-state-feedback matrix, and matrix ${\mathbf W}$ with zero 
%diagonal shows the inter-neurons
%connection weight matrix. 

In eq.(\ref{eq:matA_W_b}), the design parameter $r$, which corresponds to $b_{ii}$ in 
the "SIR" (${\theta_i}$) definition in eq.(\ref{eq:cirA}), is positive, $r>0$.  For the sake of brevity, 
we assume that the $b_i=0, \quad i=1,2, \dots, N$ in the analysis. 
%The $w_{ij}$ in eq.(\ref{eq:matA_W_b}) corresponds to $b_{ij}$ in eq.(\ref{eq:cirA}). 

It's well known that desigining the weight matrix ${\mathbf W}$ as a symmetric one yields that all 
eigenvalues are real (see e.g. \cite{Bretscher05}), 
which we assume throughout the paper for the sake of brevity of its analysis. 
%because then the implementation is simplified to a classical RC circuit design 
%%as will be shown later (in Fig. \ref{fig:SAL-USIR}). 
%In what follows, we present one of the main results of this paper: 

\vspace{0.2cm}
\emph{Proposition 1:} 
\vspace{0.2cm}

Let's assume that the linear dynamic network of eq.(\ref{eq:Diff_Linear}) with a real symmetric 
${\bf W}$ in (\ref{eq:matA_W_b}) has positive eigenvalue(s). If $r>0$ is chosen such that it's 
smaller than the maximum (positive) eigenvalue of ${\bf W}$, then 

\begin{enumerate} 

\item

the defined "SIR" (${\theta_i}$) in eq.(\ref{eq:cirA}) for any state $i$ 
asymptotically converges to the following constant as time evolves for any initial vector ${\bf x}(0)$ which is 
not completely perpendicular to the eigenvector corresponding to the largest eigenvalue of ${\bf W}$. 
\footnote{
It's easy to check in advance if the initial vector ${\bf x}(0)$ is 
completely perpendicular to the eigenvector of the maximum (positive) eigenvalue of ${\bf W}$ or not. 
If this is the case, then this can easily be overcome 
by introducing a small random number to ${\bf x}(0)$ so that it's not completely perpendicular to the 
mentioned eigenvector.
} 

\begin{equation} \label{eq:theta_const} 
%\theta_i(t > t_T) 
\theta_i^{ult} = 
\theta^{ult} =  \frac{r}{ \lambda_{max} }, \quad \quad i=1,2, \dots, N 
\end{equation}

where $\lambda_{max}$ is the maximum (positive) eigenvalue of the weight matrix ${\bf W}$. 
%and $t_T$ shows a finite time period. 

\item

there exists a finite time constant $t_T$ for a given small positive number $\epsilon_c >0$ such that 

\begin{equation} \label{eq:theta_eps} 
||{\bf \theta}^{vec}(t \geq t_T) - {\bf \theta}^{ult, vec} || < \epsilon_c 
\end{equation} 

where ${\bf \theta}^{vec}(t) = [\theta_1(t) \dots \theta_N(t)]^T$ and 
${\bf \theta}^{ult, vec} = \theta^{ult} [1 \dots 1]^T$ and $|| \cdot ||$ shows a vector norm.

\end{enumerate}

\vspace{0.2cm}
\begin{proof} 
\vspace{0.2cm}

Defining the following matrix series equation 
%(see e.g. \cite{Luenberger79}) 

\begin{eqnarray} 
{\bf e}^{( -r {\bf I+W} )t}  &  =  &  \sum_{k=0}^{\infty} \frac{(-r {\bf I+W})^k  t^k }{ k! } \label{eq:StTrMa} \\
			&  =  &  {\bf I} + \frac{( -r {\bf I+W})^1 t^1 }{ 1!} + \frac{(-r{\bf I+W})^2 t^2 }{ 2!} + \nonumber \\
			&     &   \dots + \frac{(-r{\bf I+W})^k t^k }{ k! } + \dots \label{eq:StTrMa2}
\end{eqnarray}

where $(!)$ shows the factorial, and ${\bf I}$ represents the identity matrix of right dimension, 
 it's well known that the solution of the autonomous dynamic linear system in eq.(\ref{eq:Diff_Linear}) is (see e.g. \cite{Luenberger79}):  

\begin{equation} \label{eq:solution}
{\bf x}(t) = {\bf e}^{( -r {\bf I+W} )t} {\bf x}(0)
\end{equation}

where ${\bf x}(0)$ shows the initial state vector at time zero. So,  
the state transition matrix of the linear system of eq.(\ref{eq:Diff_Linear}) is 
${\bf e}^{ ( -r {\bf I+W} ) t}$ in eq. (\ref{eq:StTrMa}) (see e.g. \cite{Luenberger79}).  

Let us first examine the powers of the matrix ($-r {\bf I} + {\bf W}$) in (\ref{eq:StTrMa}) 
in terms of matrix $r{\bf I}$ and the eigenvectors of matrix ${\bf W}$. First let's remind 
some spectral features of the symmetric real square matrices that we use in the proof later on:

It's well known that any symmetric real square matrix can be decomposed into 

\begin{equation} \label{eq:symW}
{\bf W} = \sum_{i=1}^{N} \lambda_i {\bf v}_i {\bf v}_i^T 
%= \sum_{i=1}^{N} \lambda_i {\bf V}_i 
\end{equation}

where $\{ \lambda_i \}_{i=1}^{N}$ and $\{ {\bf v}_i \}_{i=1}^{N}$ show the (real) 
eigenvalues and the corresponding orthonormal eigenvectors (see e.g. \cite{Bretscher05}), i.e., 
%and the eigenvectors $\{ {\bf v}_i \}_{i=1}^{N}$ are orthonormal (see e.g. \cite{Bretscher05}), i.e., 

\begin{equation} \label{eq:v_ortnor} 
{\bf v}_i^T {\bf v}_j = 
\left\{ 
\begin{array}{ll} 
1 & \textrm{if and only if} \quad i=j \\
0 & \textrm{otherwise,} 
\end{array}
\right.
\end{equation}

where $\quad i,j=1,2,\dots,N$. Defining the outer-product matrices of the eigenvectors  $\{ \lambda_i \}_{i=1}^{N}$ as 
${\bf V}_j = {\bf v}_i {\bf v}_i^T$ in eq. \ref{eq:symW} gives 

\begin{equation} \label{eq:symW_V}
{\bf W} = \sum_{i=1}^{N} \lambda_i {\bf V}_i
\end{equation}

The matrix ($-r{\bf I} + {\bf W}$) can be written using eq.(\ref{eq:symW}) as follows 

\begin{equation} \label{eq:W1}
-r{\bf I} + {\bf W} = -r {\bf I} + \sum_{i=1}^{N} \beta_i(1) {\bf V}_i 
\end{equation}

where $r$ is the diagonal element of matrix $r{\bf I}$, and where 
$\beta_i(1)$ is equal to 

\begin{equation} \label{eq:beta1}
\beta_i(1) = \lambda_i
\end{equation}

The matrix $(-r {\bf I + W})^2$ can be written using eq.(\ref{eq:symW})-(\ref{eq:symW_V}) as 

\begin{equation} \label{eq:W2}
(-r{\bf I} + {\bf W})^2 = r^2 {\bf I} + \sum_{i=1}^{N} \beta_i(2) {\bf V}_i
\end{equation}

where $\beta_i(2)$ is equal to 

\begin{equation} \label{eq:beta2} 
\beta_i(2) = - r \lambda_i + ( -r + \lambda_i ) \beta(1) 
\end{equation} 

Similarly, the matrix $(-r{\bf I} + {\bf W})^3$ is obtained as 

\begin{equation} \label{eq:W3} 
(-r{\bf I} + {\bf W})^3 = -r^3 {\bf I} + \sum_{i=1}^{N} \beta_i(3) {\bf V}_i 
\end{equation} 

where $\beta_i(3)$ is

\begin{equation} \label{eq:beta3} 
\beta_i(3) = r^2 \lambda_i + ( -r + \lambda_i ) \beta(2) 
\end{equation}

For $k=4$, 

\begin{equation} \label{eq:W4}
(-r{\bf I} + {\bf W})^4 = r^4 {\bf I} + \sum_{i=1}^{N} \beta_i(4) {\bf V}_i
\end{equation}

where $\beta_i(4)$ is

\begin{equation} \label{eq:beta4} 
\beta_i(4) = -r^3 \lambda_i + ( -r + \lambda_i ) \beta(3) 
\end{equation} 

So, when we continue, we observe that the $k$'th power of the matrix $(-r {\bf I + W})$ is obtained as 

\begin{equation} \label{eq:Wk}
(-r {\bf I + W})^k =(-r)^k {\bf I} + \sum_{i=1}^{N} \beta_i(k) {\bf V}_i
\end{equation}

where $\beta_i(k)$ for $k \geq 2$ is equal to 

\begin{equation} \label{eq:beta_k} 
\beta_i(k) = (-r)^{k-1} \lambda_i  +  ( -r + \lambda_i ) \beta_i(k-1)  \quad \quad k=2, 3, 4, \dots 
\end{equation} 

%In short, from eq.(\ref{eq:beta1}) and (\ref{eq:beta_k}), 

%\begin{equation} \label{eq:beta_i_k}
%\beta_i(k) = 
%\left\{ 
%\begin{array}{ll}
%\alpha \lambda_i & \textrm{for} k=1,  \\
%\alpha (1- \alpha r)^{k-1} \lambda_i + (1- \alpha r) \beta_i(k-1) + \alpha \beta_i(k-1) \lambda_i  & \textrm{for} k=2, %3, 4, \dots 
%\end{array}
%\right.
%\end{equation}

%where $r$ is the diagonal element of matrix ${\bf rI}$.  

In what follows, we summarize the findings about the auxiliary variable $\beta_i(k)$ 
from eq.(\ref{eq:beta1}) and (\ref{eq:beta_k}): 

%Using eq.(\ref{eq:beta1}) and (\ref{eq:beta_k}), the $\beta_i(k)$ is obtained as 

\begin{eqnarray}  
\beta_i(1) &  = & \lambda_i   \label{eq:betasum1} \\
\beta_i(2) &  = & -r \lambda_i + (\lambda_i - r) \lambda_i   \label{eq:betasum2} \\
\beta_i(3) &  = & r^2 \lambda_i - r \lambda_i (\lambda_i - r)  +  (\lambda_i - r)^2 \lambda_i  \label{eq:betasum3} \\ 
    	   &  \vdots  &    \label{eq:betasumdots}    \\
\beta_i(k)   & = & \sum_{m=1}^{k} \lambda_i (-r)^{k-m} (\lambda_i - r)^{m-1}		\label{eq:betasumk} \\
	  & = & \lambda_i \Big( (-r)^{k-1} + (-r)^{k-2} (\lambda_i - r)^{1} \nonumber \\
          &   & + (-r)^{k-3} (\lambda_i - r)^{2} + \dots - r (\lambda_i - r)^{k-2} \nonumber \\
          &   &  + (\lambda_i - r)^{k-1} \Big)	\label{eq:betasumk2}
\end{eqnarray} 

Defining $\zeta_i = \lambda_i / r$, we write the equation eq.(\ref{eq:betasumk2}) as follows 

%\begin{equation} \label{eq:zeta}
%\lambda_i  = \zeta_i r 
%\beta_i(k) = (-r)^{k-1} \lambda_i  +  ( -r + \lambda_i ) \beta_i(k-1)  \quad \quad i=1, 1, \dots, N 
%\end{equation} 

\begin{eqnarray}  
\beta_i(k) & = & r^k \zeta_i (-1)^{k-1} \Big( 1 - (\zeta_i-1) + (\zeta_i-1)^2 - \dots \nonumber \\
           &   &  -(-1)^{k-1}(\zeta_i-1)^{k-2} \nonumber \\
           &   &  + (-1)^{k-1} (\zeta_i-1)^{k-1} \Big) \label{eq:betazeta} 
\end{eqnarray} 

Defining the sum $S(k)$ 

\begin{eqnarray}  
S(k) &  = &  (-1)^{k-1} \Big( 1 - (\zeta_i-1) + (\zeta_i-1)^2 - \dots \nonumber \\ 
     &    &   -(-1)^{k-1}(\zeta_i-1)^{k-2} \nonumber \\
     &    &   + (-1)^{k-1} (\zeta_i-1)^{k-1} \Big) \label{eq:Series}
\end{eqnarray} 

the eq.(\ref{eq:betazeta}) is written as  

\begin{equation}  \label{eq:betazeta1}
\beta_i(k) =  r^k \zeta_i S(k) 	
\end{equation}

Summing $S(k)$ with $(\zeta_i-1)S(k)$ yields $S(k)$ as follows 

\begin{equation}  \label{eq:S(k)}
S(k) = \frac{(-1)^{k-1} + (\zeta_i-1)^k}{\zeta_i}
\end{equation}

From eq.(\ref{eq:betazeta1}) and (\ref{eq:S(k)}) 

\begin{equation}  \label{eq:betazeta2}
\beta_i(k) =  r^k \Big( (-1)^{k-1} + (\zeta_i-1)^{k} \Big) 
\end{equation}

From eq.(\ref{eq:StTrMa}), (\ref{eq:Wk}) and 
%(\ref{eq:beta_k}), 
(\ref{eq:betazeta2}), the state transition matrix is 

\begin{eqnarray} 
{\bf e}^{( -r {\bf I} + {\bf W} )t}  &  = &  \sum_{k=0}^{\infty} 
	\frac{(-r {\bf I} + {\bf W})^k  t^k)}{ k! } \label{eq:terms} \\ 
	& = &  \sum_{k=0}^{\infty} \frac{ (-r t)^k {\bf I}}{ k! }  
		+ \sum_{k=0}^{\infty} \sum_{i=1}^{N} \beta_i(k) \frac{t^k}{k!} {\bf V}_i \label{eq:terms2}
%	& = &  \sum_{k=0}^{\infty} \frac{ (-r t)^k {\bf I}}{ k! }  + 
%		\sum_{k=0}^{\infty}  \sum_{i=1}^{N} \frac{ (-\frac{\lambda_i}{r}) (-r t)^k {\bf V}_i }{ k! } +
%			\sum_{k=0}^{\infty}  \sum_{i=1}^{N} (-r+ \lambda_i) \beta_i(k-1) \frac{t^k}{k!} {\bf V}_i\label{eq:terms2}
\end{eqnarray}

The first phrase of eq.(\ref{eq:terms2}) is equal to the exponential matrix series of $-rt{\bf I}$, i.e.,  

\begin{equation} \label{eq:1stTerm}
\sum_{k=0}^{\infty} \frac{ (-r t)^k {\bf I}}{ k! } = {\bf e}^{-rt{\bf I}}
\end{equation}

The second phrase in eq.(\ref{eq:terms2}), where the sums over $i$ and $k$ are interchangable,  is obtained 
using eq.(\ref{eq:betazeta2}) as 

\begin{eqnarray} 
\sum_{i=1}^{N} \sum_{k=0}^{\infty} \beta_i(k) \frac{t^k}{k!} {\bf V}_i & = & 
	\Big( \sum_{i=1}^{N} \sum_{k=0}^{\infty} -\frac{(-r t)^k}{k!} + \nonumber \\
	 &  &  \sum_{i=1}^{N} \sum_{k=0}^{\infty} \frac{\big( r(\zeta_i -1) t \big)^k}{k!} \Big) {\bf V}_i   \label{eq:betaV} \\ 
	&  =  &  \Big( \sum_{i=1}^{N} -e^{-rt} + \sum_{i=1}^{N} e^{r(\zeta_i-1)t} \Big) {\bf V}_i \label{eq:betaV2} 
\end{eqnarray}

%We observe that the phrase in eq.(\ref{eq:1stTerm}), which is the first term of 
%the state transition matrix in eq.(\ref{eq:terms2}), as well as the first phrase in eq.(\ref{eq:betaV2})  
%exponentially goes to zero because of the term $e^{-rt}$ which exponentially goes to zero. 
%So, using eq.(\ref{eq:terms2})-(\ref{eq:betaV2}), and the definition $\lambda_i = \zeta_i r$ 

%\begin{eqnarray} 
%{\bf e}^{( -r {\bf I} + {\bf W} )t} & = & \sum_{i=1}^{N} e^{r(\zeta_i-1)t} {\bf V}_i + \epsilon_1 {\bf 1_m}, 
%	\quad \quad  t \geq t_{T_1}  \label{eq:trMzeta} \\
%	& = &  \sum_{i=1}^{N} e^{(\lambda_i-r)t} {\bf V}_i, \quad \quad  t \geq t_T  \label{eq:trMzeta2} 
%\end{eqnarray}

%In eq.(\ref{eq:trMzeta2}), if ($\lambda_i - r < 0$), in other words, if $\lambda_i < r$, then the $i$'th term 
%goes to zero exponentially. Note that $r>0$ as mentioned above. 
So, let's put the eigenvalues into two sets: Let those eigenvalues which are smaller that $r$, 
belong to set $T = \{ \lambda_{j_t} \}_{j_t=1}^{N_t}$ where $N_t$ is the length of the set; and let 
those eigenvalues which are larger than $r$ belong to set $S=\{ \lambda_{j_s} \}_{j_s=1}^{N_s}$ 
where $N_s$ is the length of the set. 
Using eqs. (\ref{eq:1stTerm})-(\ref{eq:betaV2}) and the definition $\lambda_i = \zeta_i r$, 
we write the state transition matrix in eq.(\ref{eq:terms}) as follows 

\begin{equation} \label{eq:stm_M2} 
{\bf e}^{( -r {\bf I} + {\bf W} )t}  = {\bf M}_{tp}(t) + {\bf M}_{sp}(t) 
\end{equation} 

where 

\begin{equation} \label{eq:M1} 
{\bf M}_{tp}(t) = {\bf e}^{-rt{\bf I}} - \sum_{i=1}^{N} e^{-rt} {\bf V}_i + 
		\sum_{j_t \in T} e^{(\lambda_{j_t}-r)t} {\bf V}_{j_t}
\end{equation} 

and

\begin{equation} \label{eq:M2} 
{\bf M}_{sp}(t) = \sum_{j_s \in S} e^{(\lambda_{j_s}-r)t} {\bf V}_{j_s} 
\end{equation}

We call the matrices ${\bf M}_{tp}(t)$ and ${\bf M}_{sp}(t)$ in (\ref{eq:M1}) and (\ref{eq:M2}) 
as transitory phase part and steady phase part, respectively, of the state transition matrix. 

In eq.(\ref{eq:M1}), because $r>0$ and $\lambda_{j_t}-r<0$ are finite numbers, the matrix 
${\bf M}_{tp}(t)$ exponentially vanishes (approaches to zero matrix) and 
there exists a finite time constant $t_{T_1}$ 
for a given small positive number $\epsilon_1 >0$ such that 

\begin{equation} \label{eq:M1_eps1}
||{\bf M}_{tp}(t \geq t_{T_1})|| < \epsilon_1 
\end{equation}

where $||\cdot||$ shows a matrix norm. From eq. (\ref{eq:stm_M2}) and (\ref{eq:M1_eps1}), 
the ${\bf M}_{tp}(t)$ affects only the transitory phase, and 
what shapes the steady state behavior is merely ${\bf M}_{sp}$. 
So, let's examine the steady phase solution in the following: 

The steady phase solution, denoted as ${\bf x}_{sp}(t)$ is obtained 
from eq.(\ref{eq:solution}) and (\ref{eq:stm_M2}) as 

\begin{equation} \label{eq:solutionMs} 
{\bf x}_{sp}(t) = {\bf M}_{sp}(t)  {\bf x}(0) = \sum_{js \in S} e^{(\lambda_{j_s} -r )t} {\bf V}_{j_s} {\bf x}(0)
\end{equation} 

Let's define the interference vector ${\bf J}_{sp}(t)$ as 

\begin{equation} \label{eq:J_intrf} 
{\bf J}_{sp}(t) = {\bf W} {\bf x}_{sp}(t) 
\end{equation} 

Using eq.(\ref{eq:symW}) in (\ref{eq:J_intrf}) and the orthonormal features in (\ref{eq:symW_V}) yields 

\begin{equation} \label{eq:J_intrf_b} 
{\bf J}_{sp}(t) = \sum_{j_s \in S} \lambda_{j_s} e^{(\lambda_{j_s} - r)t} {\bf V}_{j_s} {\bf x}(0)
\end{equation} 

Defining ${\bf V}_{j_s} {\bf x}(0) = {\bf u}_{j_s}$, we rewrite the eq.(\ref{eq:solutionMs}) 
and (\ref{eq:J_intrf_b}), respectively, as 

\begin{equation} \label{eq:solutionMs_u} 
{\bf x}_{sp}(t) = {\bf M}_{sp}(t)  {\bf x}(0) = \sum_{js \in S} e^{(-r + \lambda_{j_s} )t} {\bf u}_{j_s}
\end{equation} 

and

\begin{equation} \label{eq:J_intrf_u} 
{\bf J}_{sp}(t) = \sum_{j_s \in S} \lambda_{j_s} e^{(-r + \lambda_{j_s} )t} {\bf u}_{j_s} 
\end{equation}

In eq.(\ref{eq:solutionMs_u}) and (\ref{eq:J_intrf_u}), we assume that 
the ${\bf u}_{j_s} = {\bf V}_{j_s} {\bf x}(0)$ corresponding to the eigenvector of the largest eigenvalue 
is different than zero vector. This means that we assume in the analysis here that 
${\bf x}(0)$ is not completely perpendicular to the mentioned eigenvector. 
This is something easy to check in advance. If it is the case, then 
this can easily be overcome 
by introducing a small random number to ${\bf x}(0)$ so that it's not completely perpendicular to the 
mentioned eigenvector.

Taking eq.(\ref{eq:M1_eps1}) into account and 
dividing the vector ${\bf x}_{sp}(t)$ of eq.(\ref{eq:solutionMs_u}) to {\bf J}(t) of 
eq.(\ref{eq:J_intrf_u}) elementwise and comparing the outcome with the "SIR" ($\theta_i(t)$) definition 
in eq.(\ref{eq:cirA}) where $b_{ii} = r$ results in 

\begin{eqnarray} 
\frac{x_{i}^{sp}(t)}{J_{sp, i}(t)} & = & \frac{1}{r} {\theta_i}(t),  \quad  \quad t \rightarrow \infty, \quad i=1, \dots, N \label{eq:theta_xJ_i} \\ 
	 & = & \frac{ \sum_{j_s \in S} e^{(-r + \lambda_{j_s} )t} {\bf u}_{j_s}(i) }{ \sum_{j_s \in S} 
\lambda_{j_s} e^{(-r + \lambda_{j_s} )t} {\bf u}_{j_s}(i) \label{eq:theta_xJ_i2} } 
%		&  =  &  \frac{1}{r} 
%\frac{ a_{ii} x_i}{ b_i + \sum_{j = 1, j \neq i}^{N} w_{ij} x_j },  
%	\quad i=1, \dots, N
\end{eqnarray} 

where ${\bf u}_{j_s}(i)$ is the $i$'th element of the vector ${\bf u}_{j_s}$.  From the analysis above, we observe that 

\begin{enumerate}

\item  If there is only one positive eigenvalue and it's a multiple one, denoted as $\lambda_{c}$
(i.e. $\lambda_{j_s} = \lambda_{c}, \quad j_s \in S$), then it's seen from (\ref{eq:theta_xJ_i2}) that 

\begin{eqnarray} \label{eq:theta_c} 
{\theta_i}(t) \rightarrow \frac{r}{\lambda_{c}}, \quad i=1, \dots, N, \quad \quad t \geq t_{T_1}
\end{eqnarray} 

\item  Similarly, if there is only one positive eigenvalue and it's a single one, 
shown as $\lambda_{c}$, (i.e., $S=\{ \lambda_{c} \}$), then  eq.(\ref{eq:theta_c}) holds.

\item  If there are more than two different (positive) eigenvalues and the largest positive  
eigenvalue is single (not multiple), then  we observe from (\ref{eq:theta_xJ_i2}) that 
the largest (positive) eigenvalue dominates the dynamics of eq.(\ref{eq:theta_xJ_i2}) as 
time evolves because of the fact that a relatively small increase in 
the power of the exponential causes exponential increase as time evolves.  This can be seen as follows: 

Let's show the two largest (positive) eigenvalues as $\lambda_{max}$ and $\lambda_{j}$ respectively and 
the difference between them as $\Delta \lambda$. So, $\lambda_{max} = \lambda_{j} + \Delta \lambda$. 
We define the following ratio 

\begin{eqnarray} 
K(t)  & = & \frac{ e^{(\lambda_{j} - r  )t}}{ e^{(-r +  \lambda_{max} )t}} \label{eq:expDeltaL} \\
   & = & \frac{1}{ e^{ (\Delta \lambda)t } },   
    \quad \quad \quad \Delta \lambda > 0    \label{eq:expDeltaL2} 
\end{eqnarray} 

In eq.(\ref{eq:expDeltaL2}), because $\Delta \lambda > 0$ is a finite number, 
the $K(t)$ exponentially vanishes (approaches to zero), as will be depicted in 
Fig. \ref{fig:ratioDeltaL}, and there exists a finite time constant $t_{T_2}$
for a given small positive number $\epsilon_2 >0$ such that 

\begin{equation} \label{eq:Kt_eps2} 
| K( t \geq t_{T_2} )| < \epsilon_2 
\end{equation} 

Similarly, for the denominator terms, we define the following ratio 

%So, the ratio of the second largest term to the largest term of the sum in the nominator is obtained, 
%after some straighforward calculations, as 

\begin{eqnarray} \label{eq:LmaxLj} 
\frac{\lambda_{j} e^{(\lambda_{j} - r  )t} }{\lambda_{max} e^{(-r +  \lambda_{max} )t} } = 
         K(t) (1 - \frac{\Delta \lambda}{ \lambda_{j} + \Delta \lambda}) < K(t)  \\ 
	   \quad \quad \Delta \lambda, \lambda_{j} > 0  \nonumber
\end{eqnarray}

\begin{figure}[htbp]
  \begin{center}
   \epsfxsize=24.0em    % scale of the figure
\leavevmode\epsffile{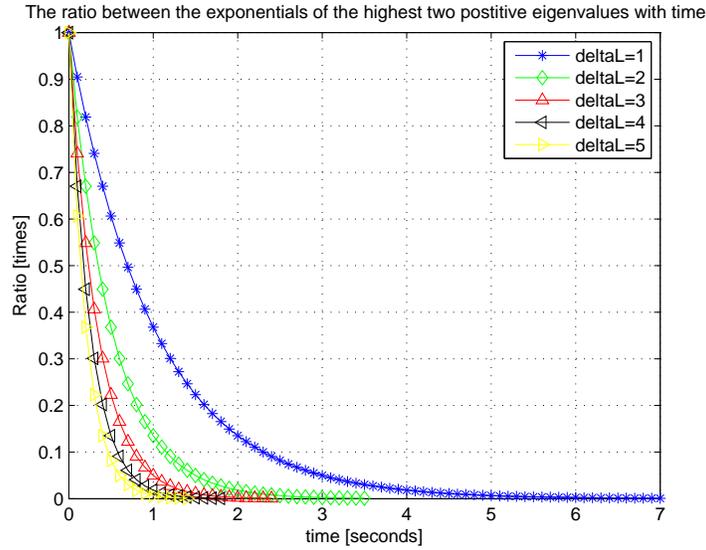}
   \vspace{-1em}        % eliminate extra empty at the bottom of the figure
  \end{center}
 \caption{ The figure shows the ratio $K$ in eq.(\ref{eq:expDeltaL}) for some different 
$\Delta \lambda$ values. }
\label{fig:ratioDeltaL}
\end{figure}

So, the $K(t)$ in eq.(\ref{eq:expDeltaL}) is obtained as a multiplier in the ratio of the 
two greatest terms of the sum in both nominator and denominator. 
We plot the ratio $K(t)$ in Fig. \ref{fig:ratioDeltaL} for some different 
$\Delta \lambda$ values.  
The Figure \ref{fig:ratioDeltaL} shows that the term related to the $\lambda_{max}$ dominates 
the sum of the nominator and that of the denominator respectively. This observation implies from 
eq.(\ref{eq:theta_xJ_i2}) that  

\begin{eqnarray} 
\frac{x_{i}^{sp}(t)}{J_{sp, i}(t)}  &  =  &  
	\frac{ \sum_{js \in S} e^{(-r + \lambda_{j_s} )t} {\bf u}_{j_s}(i) }{ \sum_{js \in S} \lambda_{j_s} e^{(-r + \lambda_{j_s} )t} {\bf u}_{j_s}(i) }  \label{eq:theta_maxL} \\
	&  \rightarrow &  
 \frac{ e^{(-r + \lambda_{max})t} }{ \lambda_{max} e^{(-r + \lambda_{max})t} } = \frac{1}{\lambda_{max}} \label{eq:theta_maxL2} 
\end{eqnarray}

\item  If the largest positive eigenvalue in case 3 above is a multiple eigenvale, then, similarly, 
the corresponding terms in the sum 
of the nominator and that of the demoninator become dominant, which implies from eq.(\ref{eq:theta_xJ_i2}) that 
$\frac{x_{i}^{sp}(t)}{J_i(t)}$ exponentially converges to $\frac{1}{\lambda_{max}}$ as time evolves. 

\end{enumerate}

Using the observations 1 to 4, and the "SIR" definition 
in eq.(\ref{eq:cirA}), we conclude from eq.(\ref{eq:theta_xJ_i}), (\ref{eq:theta_maxL}) and (\ref{eq:theta_maxL2}) that 
for any initial vector ${\bf x}(0)$ which is 
not completely perpendicular to the eigenvector corresponding to the largest eigenvalue of ${\bf W}$, 

\begin{equation} \label{eq:cirA_sp}
{\theta_i}(t \rightarrow \infty) = \frac{ r x_i^{sp}(t)}{
%\sum_{j = 1, j \neq i}^{N} w_{ij} x_j^{sp}(t) 
J_{sp, i}(t)
} \rightarrow  
 \frac{r}{\lambda_{max}},  	\quad i=1, \dots, N, 
%\quad t \geq t_T 
\end{equation}

which completes the first part of the proof.  Furthermore, from eq. (\ref{eq:M1_eps1}), (\ref{eq:Kt_eps2}) 
and (\ref{eq:cirA_sp}) 
we conclude that 
there exists a finite time constant $t_T$ for a given small positive number $\epsilon_c >0$ such that

\begin{equation} \label{eq:theta_eps_prf}
|{\bf \theta}^{vec}(t \geq t_T) - {\bf \theta}^{ult, vec} | \leq \epsilon_c 
\end{equation}

where ${\bf \theta}^{vec}(t) = [\theta_1(t) \dots \theta_N(t)]^T$ and
${\bf \theta}^{ult, vec} = \theta^{ult} [1 \dots 1]^T$, which completes the proof. 

\end{proof}

\vspace{0.2cm}
\emph{Definition:} System-Specific Ultimate SIR value: In proposition 1, we showed that the SIR in (\ref{eq:cirA}) 
for every state in the autonomous linear dynamic networks in eq.(\ref{eq:Diff_Linear}) converges to 
a constant value as time goes to infinity.  We call this converged constant value 
as "system specific ultimate SIR" and denote as ${\theta}^{ult}$: 

\begin{equation} \label{eq:theta_ult}
{\theta}^{ult} = \frac{r}{\lambda_{max}}
\end{equation}

where $r > 0$ is the design parameter and $\lambda_{max}$ 
is the maximum (positive) eigenvalue of matrix ${\bf W}$. 

%**************** 

\subsection{ Discrete-time Analysis \label{Section:analysisDiscr} } 

In this subsection, we analyse the defined ``SIR'' for the the underlying linear part of the proposed 
discrete-time autonomous network which 
is obtained by discretizing the continuous-time system of eq.(\ref{eq:Diff_Linear}) 
by using well-known Euler method: 

\begin{equation} \label{eq:Diff_Linear_discrete} 
{\mathbf x}(k+1) = \big( {\bf I} + \alpha ( -r {\bf I} + {\bf W} ) \big) {\mathbf x}(k) 
\end{equation} 

where ${\bf I}$ is the identity matrix, $r$ is a positive real number, 
$-r {\bf I}$ and ${\bf W}$ is as eq.(\ref{eq:matA_W_b}), 
${\mathbf x}(k)$ shows the state vector at step $k$, and 
$\alpha >0$ is the step size.

\vspace{0.2cm}
\emph{Proposition 2:} 
\vspace{0.2cm}

In the autonomous discrete-time linear network of eq.(\ref{eq:Diff_Linear_discrete}), 
%with a real symmetric ${\bf W}$ in (\ref{eq:matA_W_b}), 
let's assume that the spectral radius of the system  matrix 
$\big( {\bf I} + \alpha ( -r {\bf I} + {\bf W} ) \big)$ is larger than 1. 
(This assumption is equal to the assumption that 
${\bf W}$ has positive eigenvalue(s) and 
$r>0$ is chosen such that $\lambda_{max} > r$, where 
$\lambda_{max}$ is the maximum (positive) eigenvalue of ${\bf W}$).  
If $\alpha$ is chosen such that $0 < \alpha r < 1$, then 

\begin{enumerate} 

\item 

the defined "SIR" ${\theta_i}(k)$ in eq.(\ref{eq:cirA}) for any state $i$ 
asymptotically converges to the ``ultimate SIR'' constant in (\ref{eq:theta_const}) 
as time step evolves for any initial vector ${\bf x}(0)$ which is 
not completely perpendicular to the eigenvector corresponding to the largest eigenvalue of ${\bf W}$. 
%\footnote{
%It's easy to check in advance if the initial vector ${\bf x}(0)$ is 
%completely perpendicular to the eigenvector of the maximum (positive) eigenvalue of ${\bf W}$ or not. 
%If this is the case, then this can easily be overcome 
%by introducing a small random variable to ${\bf x}(0)$ so that it's not completely perpendicular to the 
%mentioned eigenvector.
%} 

%\begin{equation} \label{eq:theta_const_ds}
%\theta_i^{ult} = \theta_i^{ult} 
%(k \rightarrow \infty) =  \frac{r}{ \lambda_{max} }, \quad \quad i=1,2, \dots, N
%\end{equation} 

%where $\lambda_{max}$ is the maximum (positive) eigenvalue of the weight matrix ${\bf W}$ 
%and $k_T$ shows a finite time constant. 

\item

there exists a finite step number $k_T$ for a given small positive number $\epsilon_d >0$ such that 

\begin{equation} \label{eq:theta_eps_ds} 
||{\bf \theta}^{vec}(k \geq k_T) - {\bf \theta}^{ult, vec} || < \epsilon_d 
\end{equation} 

where ${\bf \theta}^{vec}(k) = [\theta_1(k) \dots \theta_N(k)]^T$ and 
${\bf \theta}^{ult, vec} = \theta^{ult} [1 \dots 1]^T$.  

\end{enumerate}

\vspace{0.2cm}
\begin{proof} 
\vspace{0.2cm}

From eq. (\ref{eq:Diff_Linear_discrete}), it's obtained  
%the solution of the autonomous discrete linear system in eq.(\ref{eq:Diff_Linear_discrete}) is 

\begin{equation} \label{eq:solution_ds}
{\mathbf x}(k) = \Big( {\bf I} + \alpha ( -r {\bf I} + {\bf W} ) \Big)^{k} {\mathbf x}(0) 
\end{equation}

where ${\bf x}(0)$ shows the initial state vector at step zero. Let us first examine the powers of the matrix 
$\Big( {\bf I} + \alpha ( -r {\bf I} + {\bf W} ) \Big)$ in (\ref{eq:solution_ds}) 
in terms of matrix $r{\bf I}$ and the eigenvectors of matrix ${\bf W}$: 
It's well known that any symmetric real square matrix can be decomposed into 

\begin{equation} \label{eq:symW_ds}
{\bf W} = \sum_{i=1}^{N} \lambda_i {\bf v}_i {\bf v}_i^T = \sum_{i=1}^{N} \lambda_i {\bf V}_i 
\end{equation}

where $\{ \lambda_i \}_{i=1}^{N}$ and $\{ {\bf v}_i \}_{i=1}^{N}$ show the (real) 
eigenvalues and the corresponding eigenvectors and 
the eigenvectors $\{ {\bf v}_i \}_{i=1}^{N}$ are orthonormal (see e.g. \cite{Bretscher05}), i.e., 

\begin{equation} \label{eq:v_ortnor_ds}
{\bf v}_i^T {\bf v}_j = 
\left\{ 
\begin{array}{ll}
1, & \textrm{if} \quad i=j, \quad \quad \textrm{where} \quad i,j=1,2,\dots,N \\
0, & \textrm{if} \quad i \neq j, 
\end{array}
\right.
\end{equation}

Let's define the outer-product matrices of the eigenvectors  $\{ \lambda_i \}_{i=1}^{N}$ as 
${\bf V}_j = {\bf v}_i {\bf v}_i^T, \quad i=1,2, \dots, N$; and, furthermore, the matrix ${\bf M}$ as

%from eq.(\ref{eq:v_ortnor_ds}), is equal to 

%\begin{equation} \label{eq:V_ortnorjMat_ds}
%{\bf V}_i = 
%\left\{ 
%\begin{array}{ll}
%{\bf I} & \textrm{if} \quad i=j, \quad \quad \textrm{where} \quad i,j = 1,2,\dots,N \\
%{\bf 0} & \textrm{if} \quad i \neq j, 
%\end{array}
%\right.
%\end{equation}

%where ${\bf I}$ is the identity matrix. 
%Defining matrix ${\bf M}$, 

\begin{equation} \label{eq:M}
{\bf M} = {\bf I} + \alpha ( -r {\bf I} + {\bf W} )
\end{equation}

which is obtained using  eq.(\ref{eq:symW_ds}) as 
%The matrix ${\bf M}$ can be written as follows 

\begin{equation} \label{eq:W1_ds}
{\bf M} = (1 - \alpha r) {\bf I} + \sum_{i=1}^{N} \beta_i(1) {\bf V}_i
\end{equation}

where $r > 0$, $\alpha > 0$, and where $\beta_i(1)$ is equal to 

\begin{equation} \label{eq:beta1_ds}
\beta_i(1) = \alpha \lambda_i.
\end{equation}

The matrix ${\bf M}^2$ can be written as 

\begin{equation} \label{eq:W2_ds}
{\bf M}^2 = (1 - \alpha r)^2 {\bf I} + \sum_{i=1}^{N} \beta_i(2) {\bf V}_i
\end{equation}

where $\beta_i(2)$ is equal to 

\begin{equation} \label{eq:beta2_ds} 
\beta_i(2) = \alpha (1-\alpha r) \lambda_i + (1-\alpha r + \alpha \lambda_i) \beta_i(1) 
\end{equation} 

Similarly, the matrix ${\bf M}^3$ can be written as 

\begin{equation} \label{eq:W3_ds}
{\bf M}^3 = (1 - \alpha r)^3 {\bf I} + \sum_{i=1}^{N} \beta_i(3) {\bf V}_i
\end{equation}

where $\beta_i(3)$ is equal to 

\begin{equation} \label{eq:beta3_ds}
\beta_i(3) = \alpha (1-\alpha r)^2\lambda_i + (1-\alpha r + \alpha \lambda_i) \beta_i(2)
\end{equation}

So, ${\bf M}^4$ can be written as 

\begin{equation} \label{eq:W4_ds}
{\bf M}^4 = (1 - \alpha r)^4 {\bf I} + \sum_{i=1}^{N} \beta_i(4) {\bf V}_i
\end{equation}

where $\beta_i(4)$ is equal to 

\begin{equation} \label{eq:beta4_ds}
\beta_i(4) = \alpha (1-\alpha r)^3\lambda_i + (1-\alpha r + \alpha \lambda_i) \beta_i(3)
\end{equation}

So, at step $k$, the matrix $({\bf M})^k$ is obtained as 

\begin{equation} \label{eq:W_k}
{\bf M}^k = (1 - \alpha r)^k {\bf I} + \sum_{i=1}^{N} \beta_i(k) {\bf V}_i
\end{equation}

where $\beta_i(k)$ is equal to 

\begin{equation} \label{eq:beta_i_k}
\beta_i(k) = \alpha (1-\alpha r)^{k-1} \lambda_i + ( 1 + \alpha ( \lambda_i - r) ) \beta_i(k-1)
\end{equation}

Using eq.(\ref{eq:beta1_ds}) and (\ref{eq:beta_i_k}), the $\beta_i(k)$ is obtained as 

\begin{eqnarray}  
\beta_i(1) &  = & \alpha \lambda_i   \label{eq:betasum1_ds} \\
\beta_i(2) &  = & \alpha \lambda_i \Big( (1-\alpha r) + \big( 1 + \alpha ( \lambda_i - r) \big) \Big) 
		  \label{eq:betasum2_ds} \\
\beta_i(3) &  = & \alpha \lambda_i \Big( (1-\alpha r)^2 + (1-\alpha r) \big( 1 + \alpha ( \lambda_i - r) \big) \nonumber \\
           &    & + \big( 1 + \alpha ( \lambda_i - r) \big)^2 \Big)  \label{eq:betasum3_ds} \\ 
    	   &  \vdots  &    \label{eq:betasumdots_ds}   \nonumber \\
\beta_i(k)   & = & \alpha \lambda_i \sum_{m=1}^{k} (1-\alpha r)^{k-m} \big( 1 + \alpha ( \lambda_i - r) \big)^{m-1}
	\label{eq:betasumk_ds}
\end{eqnarray} 

Defining $\lambda_i = \zeta_i (1- \alpha r)$, we obtain 

\begin{equation}
(1-\alpha r)^{k-m} \big( 1 + \alpha ( \lambda_i - r) \big)^{m-1} = (1-\alpha r)^{k-1} (1 + \alpha \zeta_i)^{m-1} 
		\label{eq:xy} 
\end{equation} 

Writing eq.(\ref{eq:xy}) in eq.(\ref{eq:betasumk_ds}) gives 

\begin{equation}  \label{eq:betazeta_ds}
\beta_i(k) = \alpha \zeta_i (1-\alpha r)^{k} S(k) 
\end{equation} 

where $S(k)$ is 

\begin{equation}  \label{eq:Series_ds}
S(k) = \sum_{m=1}^{k} (1 + \alpha \zeta_i)^{m-1} 
\end{equation}

Summing $-S(k)$ with $(1 + \alpha \zeta_i) S(k)$ yields 

\begin{equation}  \label{eq:S(k)_ds}
S(k) = \frac{ (1 + \alpha \zeta_i)^{k} - 1 }{ \alpha \zeta_i }
\end{equation}

From eq.(\ref{eq:betazeta_ds}), (\ref{eq:Series_ds}) and (\ref{eq:S(k)_ds}), we obtain  

\begin{equation}  \label{eq:betazeta1_ds}
\beta_i(k) = (1-\alpha r)^{k} (1 + \alpha \zeta_i)^{k} - (1-\alpha r)^{k}
\end{equation}

Using the definition $\zeta_i = \lambda_i / (1- \alpha r)$ in  eq.(\ref{eq:betazeta1_ds}) gives 

\begin{equation}  \label{eq:betazeta2_ds} 
\beta_i(k) = \big( 1 + \alpha ( \lambda_i - r) \big)^{k} - (1-\alpha r)^{k} 
\end{equation} 

From eq.(\ref{eq:W_k}) and eq.(\ref{eq:betazeta2_ds}), 

\begin{equation} \label{eq:W_kfromZeta}
{\bf M}^k = (1 - \alpha r)^k {\bf I} 
		+ \sum_{i=1}^{N} \big( 1 + \alpha ( \lambda_i - r) \big)^{k} {\bf V}_i 
			- \sum_{i=1}^{N} (1-\alpha r)^{k} {\bf V}_i 
\end{equation}

Let's put the $N$ eigenvalues of matrix ${\bf W}$ into two groups as follows: 
Let those eigenvalues which are smaller that $r$, 
belong to set $T = \{ \lambda_{j_t} \}_{j_t=1}^{N_t}$ where $N_t$ is the length of the set; and let 
those eigenvalues which are larger than $r$ belong to set $S=\{ \lambda_{j_s} \}_{j_s=1}^{N_s}$ 
where $N_s$ is the length of the set. We write the matrix ${\bf M}^k$ in eq.(\ref{eq:W_kfromZeta}) using 
this eigenvalue grouping 

\begin{equation} \label{eq:stm_M} 
{\mathbf M}^k  = {\bf M}_{tp}(k) + {\bf M}_{sp}(k) 
\end{equation} 

where 

\begin{eqnarray} 
{\bf M}_{tp}(k) & = & (1 - \alpha r)^k {\bf I} - \sum_{i=1}^{N} (1-\alpha r)^{k} {\bf V}_i \nonumber \\ 
		&   &  + \sum_{j_t \in T} \big( 1 + \alpha ( \lambda_{j_t} - r) \big)^{k} {\bf V}_{j_t} \label{eq:M1_ds} 
\end{eqnarray} 

and 

\begin{equation} \label{eq:M2_ds} 
{\bf M}_{sp}(k) = \sum_{j_s \in S} \big( 1 + \alpha ( \lambda_{j_s} - r) \big)^{k} {\bf V}_{j_s} 
\end{equation}

We call the matrices ${\bf M}_{tp}(k)$ and ${\bf M}_{sp}(k)$ in (\ref{eq:M1_ds}) and (\ref{eq:M2_ds}) 
as transitory phase part and steady phase part, respectively, of the matrix ${\mathbf M}^k$. 

It's observed from eq.(\ref{eq:M1_ds}) that the ${\bf M}_{tp}(k)$ converges to zero matrix as time step number evolves 
because relatively small step number $\alpha >0$ is chosen such that $(1-\alpha r) < 1$ and 
$1 + \alpha ( \lambda_{j_t} - r) < 1$. Therefore, 
there exists a finite time step number $k_{T_1}$ for a given small positive number $\epsilon_3 >0$ such that

\begin{equation} \label{eq:nMtp_eps3} 
|| {\bf M}_{tp}(k \geq k_{T_1}) || < \epsilon_3 
\end{equation} 

Thus, (from eq.(\ref{eq:M}) and (\ref{eq:stm_M})), 
what shapes the steady state behavior of the system in eq.(\ref{eq:solution_ds}) is merely the 
${\bf M}_{sp}(k)$ in eq.(\ref{eq:M2_ds} ). So, the steady phase solution is obtained from eqs.(\ref{eq:solution_ds}), 
(\ref{eq:M}) and (\ref{eq:stm_M})-(\ref{eq:M2_ds}) using the above observations as follows 

\begin{eqnarray}  
{\bf x}_{sp}(k) & = & {\bf M}_{sp}(k)  {\bf x}(0) \label{eq:solutionMs_ds} \\ 
	&  =  &	 \sum_{j_s \in S} \big( 1 + \alpha ( \lambda_{j_s} - r) \big)^{k} {\bf V}_{j_s} {\bf x}(0) 
%\quad \quad k \geq k_T 
\label{eq:solutionMs2_ds}  
\end{eqnarray}

Let's define the interference vector, ${\bf J}_{sp}(k)$ as 

\begin{equation} \label{eq:J_intrf_ds} 
{\bf J}_{sp}(k) = {\bf W} {\bf x}_{sp}(k) 
\end{equation} 

Using eq.(\ref{eq:symW_ds}) in (\ref{eq:J_intrf_ds}) and the orthonormal features in (\ref{eq:v_ortnor_ds}) yields 

\begin{equation} \label{eq:J_intrf_b_ds} 
{\bf J}_{sp}(k) = 
\sum_{j_s \in S} \lambda_{j_s} \big( 1 + \alpha ( \lambda_{j_s} - r) \big)^{k} {\bf V}_{j_s} {\bf x}(0)
\end{equation}

First defining ${\bf V}_{j} {\bf x}(0) = {\bf u}_{j}$, 
and $\xi = \frac{\alpha}{1-\alpha r}$, 
then dividing vector ${\bf x}_{sp}(k)$ of eq.(\ref{eq:solutionMs2_ds}) to ${\bf J}_{sp}(k)$ of 
eq.(\ref{eq:J_intrf_b_ds}) elementwise and comparing the outcome with the "SIR" definition in 
eq.(\ref{eq:cirA}) results in 

\begin{eqnarray}  
\frac{x_{sp, i}(k)}{J_{sp, i}(k)} & = & \frac{1}{r} {\theta_i}(k),  
		\quad  \quad \quad \quad \quad i=1, \dots, N \label{eq:theta_xJ_i_ds} \\ 
	 & = & 
	\frac{ \sum_{j_s \in S} ( 1 + \xi \lambda_{j_s} )^{k} u_{j_s, i} }{ \sum_{j_s \in S} \lambda_{j_s} ( 1 + \xi \lambda_{j_s} )^{k} u_{j_s, i} }
	\label{eq:theta_xJ_i2_ds} 
\end{eqnarray}

In eq.(\ref{eq:theta_xJ_i2_ds}), we assume that 
the ${\bf u}_{j} = {\bf V}_{j} {\bf x}(0)$ which corresponds to the eigenvector of the largest positive eigenvalue 
is different than zero vector. This means that we assume in the analysis here that 
${\bf x}(0)$ is not completely perpendicular to the mentioned eigenvector. 
%This is something easy to check in advance. If it is the case, then 
%this can easily be overcome 
%by introducing a small random number to ${\bf x}(0)$ so that it's not completely perpendicular to the 
%mentioned eigenvector. 

From the analysis above, we observe that 

\begin{enumerate}

\item If there is only one positive eigenvalue which is greater than $r$ and it's a multiple 
one, denoted as $\lambda_{b}$, then it's seen from (\ref{eq:theta_xJ_i2_ds}) that 

\begin{eqnarray} \label{eq:theta_c_ds} 
{\theta_{i}}(k) = \frac{r}{\lambda_{b}}, \quad k \rightarrow \infty, \quad i=1, \dots, N 
\end{eqnarray} 

\item  Similarly, if there is only one positive eigenvalue which is larger than $r$ 
and it's a single one, 
shown as $\lambda_{b}$, then  eq.(\ref{eq:theta_c_ds}) holds.

\item  If there are more than two different (positive) eigenvalues and the largest positive  
eigenvalue is single (not multiple), then  we see from (\ref{eq:theta_xJ_i_ds}) that 
the term related to the largest (positive) eigenvalue dominates the sum of the nominator. 
Same observation is valid for the sum of the denominator. 
This is because a relatively small increase in $\lambda_{j}$ 
causes exponential increase as time step evolves, which is shown in the following: 
Let's show the two largest (positive) eigenvalues as $\lambda_{max}$ and $\lambda_{j}$ respectively and 
the difference between them as $\Delta \lambda$. So, $\lambda_{max} = \lambda_{j} + \Delta \lambda$. 
Let's define the following ratio between the terms related to the two highest eigenvalues 
in the nominator  

\begin{equation} \label{eq:expDeltaL_ds}
K_n(k)  = \frac{ (1+\xi \lambda_{j})^{k} }{(1+\xi (\lambda_{j} + \Delta \lambda) )^{k}} 
\end{equation} 

where 

\begin{equation} \label{eq:xi} 
\xi = \frac{\alpha}{1-\alpha r}. 
\end{equation} 

In eq.(\ref{eq:expDeltaL_ds}), because $\Delta \lambda > 0$, 
there exists a finite time step number $k_{T_2}$
for a given small positive number $\epsilon_3 >0$ such that

\begin{equation} \label{eq:Kn_eps3}
|K_n(k)| < \epsilon_3
\end{equation}

Similarly, let's define the ratio between the terms related to the two highest eigenvalues 
in the denominator as 

\begin{equation} \label{eq:expDeltaL_d}
K_d(k) = \frac{ \lambda_j (1+\xi \lambda_{j})^{k} }{ (\lambda_{j}+\Delta \lambda) (1+\xi (\lambda_{j} + \Delta \lambda) )^{k} }
\end{equation} 

From eq.(\ref{eq:expDeltaL_ds}) and (\ref{eq:expDeltaL_d}), because $\Delta \lambda > 0$ and 
$\frac{\lambda_{j}}{\lambda_{j} + \Delta \lambda} < 1$, 

\begin{equation} \label{eq:K_dn} 
K_d(k) < K_n(k).
\end{equation} 

and there exists a finite time step number $k_{T_2}$ for a given small positive number $\epsilon_4 >0$ such that

\begin{equation} \label{eq:K_n_eps4}
| K_n( k \geq k_{T_2} ) | < \epsilon_4.
\end{equation}

\begin{figure}[htbp]
  \begin{center}
   \epsfxsize=24.0em    % scale of the figure
\leavevmode\epsffile{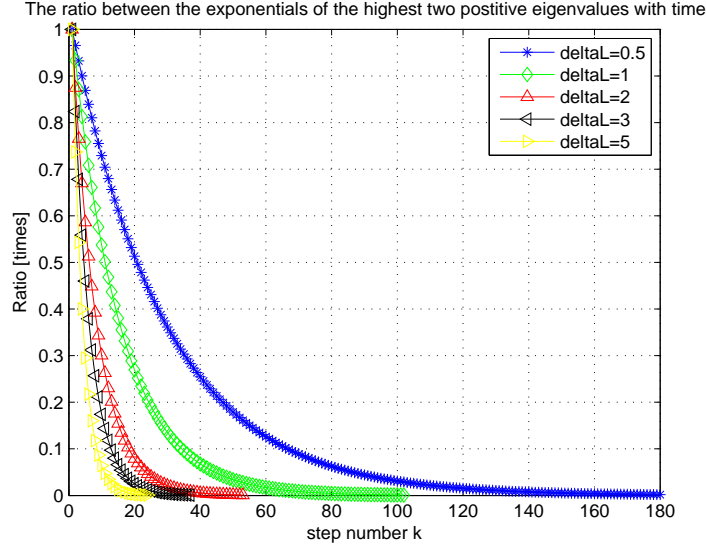}
   \vspace{-1em}        % eliminate extra empty at the bottom of the figure
  \end{center}
 \caption{ The figure shows the ratio $K_n$ in eq.(\ref{eq:expDeltaL_ds}) for some different
$\Delta \lambda$ values ($\lambda=5, \xi=0.11$). }
\label{fig:ratioDeltaL_ds}
\end{figure}

We plot the ratio $K_n(k)$ in Fig. \ref{fig:ratioDeltaL_ds} for some different $\Delta \lambda$ values and 
for a typical $\xi$ value.  
The Figure \ref{fig:ratioDeltaL_ds} and eq.(\ref{eq:K_dn}) implies that 
the terms related to the $\lambda_{max}$ dominate 
the sum of the nominator and that of the denominator respectively. 
So, from eq.(\ref{eq:theta_xJ_i2_ds}) and (\ref{eq:xi}), 

\begin{eqnarray} 
\frac{x_{sp, i}(k)}{J_{sp, i}(k)}  &  = &  
\frac{ \sum_{j_s \in S} ( 1 + \xi \lambda_{j_s} )^{k} u_{j_s, i} }{ \sum_{j_s \in S} \lambda_{j_s} ( 1 + \xi \lambda_{j_s} )^{k} u_{j_s, i} }  \nonumber \\
      & \rightarrow & 
\frac{ ( 1 + \xi \lambda_{max} )^{k} }{ \lambda_{max} ( 1 + \xi \lambda_{max} )^{k} } = \frac{1}{\lambda_{max}}, 
%\quad k \geq k_T  \label{eq:theta_maxL_ds} 
\end{eqnarray}

\item  If the largest positive eigenvalue in case 3 above is a multiple eigenvale, then, similarly, 
the corresponding terms in the sum 
of the nominator and that of the demoninator become dominant, which implies from eq.(\ref{eq:theta_xJ_i2_ds}), 
(\ref{eq:expDeltaL_ds}) and (\ref{eq:expDeltaL_d}) that 
$\frac{x_{sp, i}(k)}{J_{sp, i}(k)}$ converges to $\frac{1}{\lambda_{max}}$ as step number increases. 

\end{enumerate}

Using the observations 1 to 4, eq.(\ref{eq:nMtp_eps3}), the "SIR" definition 
in eq.(\ref{eq:cirA}), eq.(\ref{eq:theta_xJ_i_ds}) and (\ref{eq:theta_xJ_i2_ds}), 
we conclude that 

\begin{equation} \label{eq:cirA_sp_ds} 
{\theta_i}(k \rightarrow \infty) = \frac{ r x_{sp, i}(k) }
{J_{sp, i}(k)} 
%{ \sum_{j = 1, j \neq i}^{N} w_{ij} x_{sp, j}(k) } = 
\rightarrow 
 	\frac{r}{\lambda_{max}}
%\quad k \geq k_T 
\end{equation} 

where $i=1, \dots, N,$ and $\lambda_{max}$ is the largest (positive) eigenvalue of the matrix {\bf W}, 
which completes the first part of the proof. Furthermore, from eq. (\ref{eq:nMtp_eps3}), (\ref{eq:K_n_eps4})
and (\ref{eq:cirA_sp_ds}), we conclude that
there exists a finite time constant $k_T$ for a given small positive number $\epsilon_d >0$ such that

\begin{equation} \label{eq:theta_eps_prf_ds}
||{\bf \theta}^{vec}(k \geq k_T) - {\bf \theta}^{ult, vec} || < \epsilon_d
\end{equation}

where ${\bf \theta}^{vec}(k) = [\theta_1(k) \dots \theta_N(k)]^T$ and
${\bf \theta}^{ult, vec} = \theta^{ult} [1 \dots 1]^T$, which completes the proof.

\end{proof} 

%********

\section{Stabilized SIR system}
\label{Section:proposedNet} 

Do the results of the ultimate SIR analysis in Section \ref{Section:USIR} above 
have any practical meanings? Our answer is yes. In this section, we propose 
two Hopfield-like networks in continuous and discrete-time domain respectively 
where the ``system-specific ultimate SIR'' 
is used to stabilize the system.

\vspace{0.2cm}
\subsection{ Continuous-time SALU-''SIR'' \label{Section:proposedNet_cont} }
\vspace{0.2cm}

Defining the following $g(a)$ function,

\begin{eqnarray}
g(a) =
\left\{
\begin{array}{ll}
1 & \textrm{if} \quad |a| \geq \epsilon, \\
0 & \textrm{otherwise}  \label{eq:gx_fn}
\end{array}
\right.
\end{eqnarray}

we propose the following dynamic network, 

\begin{eqnarray} 
\dot{ {\mathbf x}} & = & ( -r{\mathbf I} + {\mathbf W} ) 
	{\mathbf x}  g ( || {\bf \theta}^{vec}(t) - {\bf \theta}^{ult, vec} || ) \label{eq:SAL-USIR_x} \\
	{\mathbf y} & = & sign( {\mathbf x} )  \label{eq:SAL-USIR_y} 
\end{eqnarray}

where ${\bf \theta}^{vec}(t) = [\theta_1(t) \dots \theta_N(t)]^T$ and
${\bf \theta}^{ult, vec} = \theta^{ult} [1 \dots 1]^T$, 
and ${\mathbf y}$ is the output of the network. 

%and 

%\begin{equation} \label{eq:delta_fn}
%{\bf \delta} ( {\bf \theta}(t) - {\bf \theta}^{ult} ) = 
%\left\{ 
%\begin{array}{ll}
%0 & \textrm{if and only if} \quad {\bf \theta} = {\bf \theta}^{ult}, \\
%1 & \textrm{otherwise} 
%\end{array}
%\right.
%\end{equation} 

We call the network in eqs.(\ref{eq:gx_fn})-(\ref{eq:SAL-USIR_y}) 
as Stabilized Autonomous Linear Networks by Ultimate ``SIR'' (SAL-U"SIR").

\begin{figure}[htbp]
  \begin{center}
   \epsfxsize=24.0em    % scale of the figure
\leavevmode\epsffile{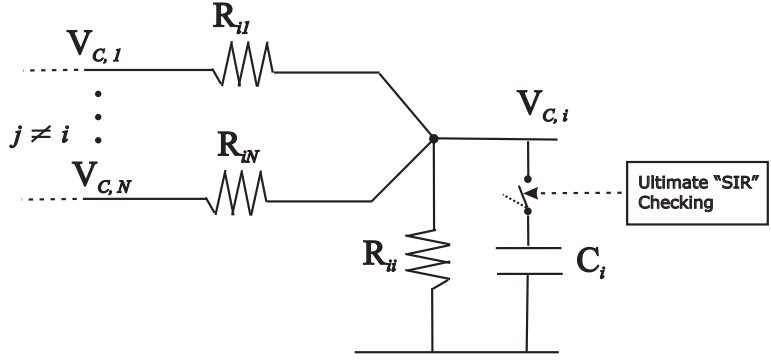}
   \vspace{-1em}        % eliminate extra empty at the bottom of the figure
  \end{center}
 \caption{ The figure depicts a sketch of the proposed network 
for numerical implementation purposes, omitting the considerations on the circuit, 
called Stabilized Autonomous Linear Networks with Ultimate ``SIR'' (SAL-U"SIR"). }
\label{fig:SAL-USIR} 
\end{figure}

\vspace{0.2cm}
\emph{Proposition 3:}
\vspace{0.2cm}

The proposed dynamic network of eqs.(\ref{eq:SAL-USIR_x})-(\ref{eq:SAL-USIR_y}) 
with the weight matrix $(-r{\bf I} + {\bf W})$ as defined in 
eq. (\ref{eq:matA_W_b}) is stable for any initial condition ${\bf x}(0)$.
% which is 
%not completely perpendicular to the eigenvector corresponding to the largest eigenvalue of ${\bf W}$. 
%%\footnote{ 
%%This condition comes from the analysis of proposition 1. 
%%} 

\vspace{0.2cm}
\begin{proof} 
\vspace{0.2cm}

If all the eigenvalues of the symmetrix matrix $(-r{\bf I} + {\bf W})$ as defined in 
eq. (\ref{eq:matA_W_b}) are negative, then it's well known from linear systems theory that 
the states go to zero exponentially for any initial vector ${\bf x}(0)$ (see e.g. \cite{Luenberger79}). 

Otherwise, if there exists positive eigenvalue(s) of $(-r{\bf I} + {\bf W})$, which is the case in our design, 
then the proposition 1 above proves for the underlying linear system of (\ref{eq:SAL-USIR_x}) that 
i) the defined SIR in eq.(\ref{eq:cirA}) for any state $i$  ($x_i(t), \quad 1, 2, \dots, N, $) 
converges, as time evolves, to the constant system-specific ultimate SIR value in eq.(\ref{eq:theta_const}) 
for any initial vector ${\bf x}(0)$, and 
ii) there exists a finite time constant $t_T$ for a given small positive number $\epsilon >0$ such that
$||{\bf \theta}^{vec}(t \geq t_T) - {\bf \theta}^{ult, vec} || < \epsilon$. 
So, the function $g(\cdot)$ stabilizes the system within the $t_T$ seconds, i.e., once
$||{\bf \theta}^{vec}(t) - {\bf \theta}^{ult, vec} || < \epsilon$ is met. 
So, the system is stable. 

\end{proof} 

\vspace{0.2cm} 

From the analysis above for symmetric $\mathbf{ W }$ and positive $r \mathbf{ I }$, we observe that  

\begin{enumerate}

\item 
The SAL-U"SIR" does not show oscilatory behaviour because all eigenvalues are real (i.e., no imaginary part), 
and at least one eigenvalue is positive, which is assured by choosing the matrix $r \mathbf{ I }$ accordingly. 
%(There exists only positive and negative eigenvalues). 

%\item 
%The $\theta_i$ converges to $\theta_i^{ult}=constant$ for any initial condition. 

\item 
The transition phase of the "unstable" linear network is shaped by the initial state vector 
and the phase space characteristics formed by the eigenvectors. 
The network is stabilized by the function $g(\cdot)$ 
once the network has passed the transition phase. 
The output of the network then is formed taking the sign of the converged states. 
Whether the state converges to a plus or minus value is dictated by the phase space shaped by the 
eigenvectors of $\mathbf{ W }$ and the initial vector $\mathbf{ x }(0)$.
\end{enumerate}

From observation 1 and 2 above, we see that choosing the 
positive matrix $r \mathbf{ I }$ such that the matrix 
$(-r{\bf I} + {\bf W})$ has positive eigenvalues makes the proposed SAL-U"SIR" exhibit 
similar features as the Hopfield Network does. The computer simulations in section \ref{Section:SimuResults} 
shows the performance of the proposed network as compared to the Hopfield Network 
in some simple associative memory systems examples. 

%as shown in the simulation results in section \ref{Section:SimuResults}. 

As far as the possible circuital implementations of the proposed network is concerned, 
a further research is needed on especially how the nonlinear function $g(\cdot)$ could be implemented 
on the circuit. In this paper, we mainly focus on the analysis part, and the circuital implementation part 
is left as a future research topic. However, 
in order to give an insight, in what follows, we propose the following simplified network 
for numerical implementation purposes:  

It's well known that a linear dynamic network like (\ref{eq:Diff_Linear}) 
%in (\ref{eq:Diff_Linear_trad}) 
can be implemented by RC (Resistance-Capacitance) circuits.  Corresponding RC dynamic network is 
given as follows: 

\begin{equation} \label{eq:RC_dyn}
\left[
\begin{array}{c}
C \dot{ V }_{C_1} \\
C \dot{ V }_{C_2}  \\
\vdots \\
C \dot{ V }_{C_N} 
\end{array}
\right]
=  
\left[
\begin{array}{c c c c}
R_{11}  &  R_{12}   & \ldots  &  R_{1N} \\
R_{21}     &   0 & \ldots  &  R_{2N} \\
\vdots &     & \ddots  &  \vdots \\
R_{N1}    &   R_{N2}   & \ldots  &  0
\end{array}
\right]
\left[
\begin{array}{c}
V_{C_1} \\
V_{C_2} \\
\vdots \\
V_{C_N}
\end{array}
\right]
\end{equation}

where $C$ represents the capacitance, $V_{C_i}$ shows the voltage of the capacitance $C_i$, which is 
the state $i$ of the network and $R_{ij}$ is the resistance. 
From eq.(\ref{eq:Diff_Linear_trad}), (\ref{eq:matA_W_b}) and (\ref{eq:RC_dyn}), 

\begin{eqnarray} 
R_{ij}  & = & \frac{1}{C} w_{ij}, \quad \quad i,j = 1, 2, \dots , N \quad \textrm{and} \quad i \neq j, \label{eq:RijWij} \\ 
R_{ii}  & = & r 
%\quad \quad \quad i = 1, 2, \dots , N  
\label{eq:RijWij2} 
\end{eqnarray} 

So, we sketch a simplified numerical implementation of the proposed SAL-U''SIR'' in Fig. \ref{fig:SAL-USIR}, 
omitting the considerations on circuits, where the function $g(\cdot)$ is represented 
by a switch (``ultimate SIR checking'').

%%%%%%%%%%%%%%%%

\vspace{0.2cm}
\subsection{ Discrete-time SALU-''SIR'' \label{Section:proposedNet_cont_ds} }
\vspace{0.2cm}

The proposed autonomous network in discrete-time, 
called DSAL-USIR (Discrete Stabilized Autonomous Linear networks by Ultimate ``SIR'')
is given as follows 

\begin{eqnarray} 
{\mathbf x}(k+1) & = & \Big( {\bf I} + \alpha ( -r {\bf I} + {\bf W} ) \Big) {\mathbf x}(k) 
                 g ( || {\bf \theta}^{vec}(t) - {\bf \theta}^{ult, vec} || )  \label{eq:SAL-USIR1} \\
	{\mathbf y}(k) & = & sign( {\mathbf x}(k) )    \label{eq:SAL-USIR1n} 
\end{eqnarray}

where ${\bf W}$ is defined by eq.(\ref{eq:matA_W_b}), the function $g(\cdot)$ is defined by eq.(\ref{eq:gx_fn}), 
$\alpha >0$ is step size, ${\bf I}$ is identity matrix and $r>0$ as in  eq.(\ref{eq:Diff_Linear_discrete}), 
${\bf \theta}^{vec}(t) = [\theta_1(t) \dots \theta_N(t)]^T$ and 
${\bf \theta}^{ult, vec} = \theta^{ult} [1 \dots 1]^T$, 
and ${\mathbf y}(k)$ is the output of the network. 

\vspace{0.2cm}
\emph{Proposition 4:} 
\vspace{0.2cm}

The proposed discrete-time networks, DSAL-U''SIR'', in eq.(\ref{eq:SAL-USIR1}) and (\ref{eq:SAL-USIR1n}) 
is stable for any initial vector ${\bf x}(0)$. 
%which is 
%not completely perpendicular to the eigenvector corresponding to the largest eigenvalue of ${\bf W}$.
%\footnote{ 
%See the footnote of proposition 1.  
%} 

\vspace{0.2cm}  

\begin{proof} 

If the spectral radius of the system  matrix
$\big( {\bf I} + \alpha ( -r {\bf I} + {\bf W} ) \big)$ is smaller than 1, then 
it's well known from the discrete-time linear systems theory that
the states go to zero exponentially for any initial vector ${\bf x}(0)$ (see e.g. \cite{Luenberger79}).

If, on the other hand, the spectral radius is larger than 1, which is the case in our design, 
then the proposition 2 above proves for the underlying linear system (\ref{eq:Diff_Linear_discrete}) that 
i) the defined ``SIR'' ${\theta_i}(k)$ in eq.(\ref{eq:cirA}) for any state $i$
asymptotically converges to the ``system-specific ultimate SIR'' constant in (\ref{eq:theta_const})
as time step evolves for any initial vector ${\bf x}(0)$ 
ii) there exists a finite step number $k_T$ for a given small positive number $\epsilon_d >0$ such that
$||{\bf \theta}^{vec}(k \geq k_T) - {\bf \theta}^{ult, vec} || < \epsilon_d$.
So, the function $g(\cdot)$ stabilizes the system within the $k_T$ steps, i.e., once
$|| {\bf \theta}^{vec}(t) - {\bf \theta}^{ult, vec} || < \epsilon_d$ is met.
So, the system is stable.

\end{proof}

As far as the design of weight matrix $\mathbf{ W }$ and $r$ is concerned, we propose to use 
the following method which is based on the well known Hebb-learning rule \cite{Hebb49}.

\subsection{ Outer products based network design  \label{Subsection:HebbBasedLearning} }

Let's assume that $L$ desired prototype vectors, $\{ \mathbf{ d }_s \}_{s=1}^{L}$,
are given from  $(-1, +1)^N$. The proposed method is based on well-known Hebb-learning \cite{Hebb49} as follows:

Step 1: Calculate the sum of outer products of the prototype vectors (Hebb Rule, \cite{Hebb49})

\begin{eqnarray} \label{eq:HebbQd}
\mathbf{ Q } = \sum_{s=1}^{L} \mathbf{ d }_s  \mathbf{ d }_s^T
\end{eqnarray}

Step 2: Determine the diagonal matrix $r \bf{I}$ and $\bf{W}$ as follows:

\begin{equation} \label{eq:rfromHebb}
r = q_{ii} + \rho
\end{equation}

where $\rho$ is a real number and

\begin{equation} \label{eq:WfromHebb}
w_{ij} =
\left\{
\begin{array}{ll}
0 & \textrm{if} \quad i = j, \\
q_{ij} & \textrm{if} \quad i \neq j
\end{array}
\right.   \quad \quad i,j=1, \dots, N
\end{equation}

where $q_{ij}$ shows the entries of matrix $\mathbf{ Q }$, $N$ is the dimension of the vector
$\mathbf{ x }(t)$ and $L$ is the number of the prototype vectors ($N > L > 0$).
From (\ref{eq:HebbQd}), $q_{ii} = L$ in eq.(\ref{eq:rfromHebb})
since $\{ \mathbf{ d }_s \}$ is from $(-1, +1)^N$.

If the desired prototype vectors are orthogonal, then it can be shown, using the steps of the proofs of 
prepositions 1 and 3 for continuous and discrete-time respectively, that 
the ``system specific ultimate SIR'' be $\theta^{ult} =  \frac{r}{ N-L }$.

\section{Simulation Results  \label{Section:SimuResults} }

We take the same examples as in \cite{Uykan08a} for comparison reasons and 
for the sake of brevity. 

In this section, we apply the 
the proposed networks SAL-U''SIR'' and DSAL-U''SIR'', in continuous and discrete-time respectively, 
to associate memory systems design, and present their simulation results as compared to those of 
corresponding Hopfield Networks. 
The weight matrices of the proposed networks and the Hopfield Networks 
are designed by the outer-products (Hebb learning \cite{Hebb49}) learning rule 
described in Section \ref{Subsection:HebbBasedLearning}. 

\subsection{Continuous-time examples}

In this section, we present two examples, one with 8 neurons and one with 16 neurons, in Example 1 and 2 respectively. 

The proposed DSALU-SIR network is given by eqs. (\ref{eq:SAL-USIR_x}) and (\ref{eq:SAL-USIR_y}). 
The Hopfield Network \cite{Hopfield85}, used as the reference network, is given by 

\begin{eqnarray} \label{eq:HopfieldNN} 
\dot{ {\mathbf x}}  &  = &  -r {\bf x} +  {\bf W} {\mathbf f}( {\bf x}(t) ) + {\bf b} \\
 	{\mathbf y}(t) &  = &  {\bf f} ({\bf x}(t))
\end{eqnarray} 

where ${\bf W}$ is the weight matrix and ${\bf x}(t)$ is the state at time $t$, 
${\bf b} = {\bf 0}$, 
${\bf f} ({\bf x}) = [f(x_1)  f(x_2)  \dots f(x_N) ]^T$, 
the $f(\cdot)$ is a sigmoid function, 
i.e., $f(x_i) = 1 - \frac{1}{1 + exp(-\sigma x_i)}$, where $\sigma>0$.

\vspace{0.2cm}
\emph{Example 1:}
\vspace{0.2cm}
 
This example is taken from example 1 in \cite{Uykan08a}.  In the design, $\sigma=2$ and $\rho$ is chosen as -2, and $r=1$. 
The desired prototype vectors are given in the raws of matrix ${\mathbf D}$ as follows, 

\begin{equation} \label{eq:ex1_D}
{\mathbf D} =
\left[
\begin{array}{c c c c c c c c}
1   &   1   & 1  &  1  & -1  &  -1  & -1  & -1  \\
1   &   1   & -1  &  -1  & 1  &  1  & -1  & -1  \\
1   &   -1  &  1  &  -1  & 1  &  -1  & 1  & -1 
\end{array}
\right]
\end{equation}

The weight matrix, using the design rule in Section \ref{Subsection:HebbBasedLearning}, is obtained as 

\begin{eqnarray} 
r {\mathbf I} = {\mathbf I}, \quad \quad \textrm{and} \quad \quad \quad \quad \quad \quad  \\
{\mathbf W} =
\left[
\begin{array}{c c c c c c c c}
0   &   1   &  1  &  -1  &  1  &  -1 & -1  & -3 \\
1   &   0   & -1  &   1  & -1  &  1  & -3  & -1 \\
1   &   -1  &  0  &   1  & -1  &  -3 & 1  & -1 \\
-1  &   1   & 1   &  0   & -3  &  -1 & -1 &  1 \\
1   &   -1  & -1  &  -3  & 0  &  1   &  1  & -1 \\
-1  &   1   &  -3 &  -1  & 1  &  0   & -1  & 1 \\
-1  &   -3  & 1   &  -1  &  1 &  -1  &  0  & 1 \\
-3  &   -1  & -1  &   1  & -1 &  1  &  1  & 0 
\end{array}
\right],  
\label{eq:matA_W_b_ex1}
%\quad \quad
%{\mathbf b} = {\mathbf 0}
\end{eqnarray}

%where ${\mathbf I}$ shows the identity matrix of dimension $N$ by $N$.

\begin{figure}[htbp]
  \begin{center}
   \epsfxsize=24.0em    % scale of the figure
\leavevmode\epsffile{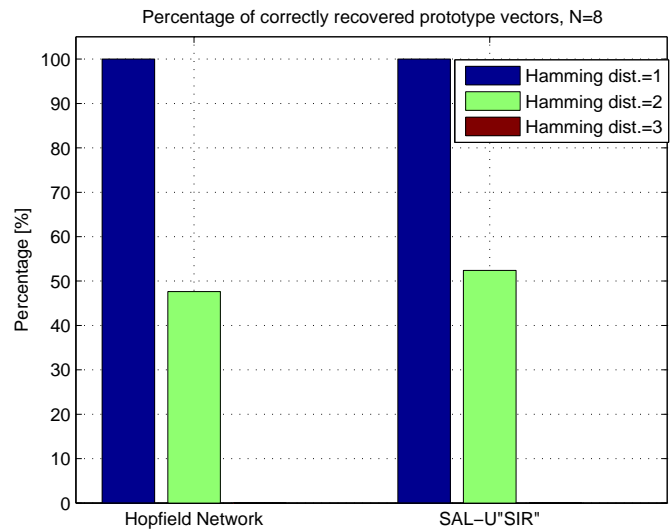}
   \vspace{-1em}        % eliminate extra empty at the bottom of the figure
  \end{center}
 \caption{ The figure
shows percentage of correctly recovered desired patterns for
all possible initial conditions in example 1 for the proposed SALU-''SIR'' 
as compared to traditional Hopfield network with 8 neurons. }
\label{fig:SALUSIR_ex1_percentage} 
\end{figure}

The Figure \ref{fig:SALUSIR_ex1_percentage} shows the percentages of correctly recovered desired patterns for 
all possible initial conditions $\mathbf{ x }(t=0) \in (-1,+1)^N$, in the proposed SALU-"SIR" 
as compared to traditional Hopfield network.  

Let $m_d$ show the number of prototype vectors and $C(N,K)$, (such that $N \geq K \geq 0$), represent the 
combination $N, K$, which is equal to $C(N,K)=\frac{N!}{(N-K)! K!}$, where $!$ shows factorial. 
In our simulation, the prototype vectors are from $(-1,1)^N$ as seen above. For initial conditions,  
we alter the sign of $K$ states where $K$=0, 1, 2, 3 and 4, which means the initial condition 
is within $K$-Hamming distance from the corresponding prototype vector. 
So, the total number of different possible combinations for the initial conditions for this example is 
24, 84 and 168 for 1, 2 and 3-Hamming distance cases respectively, which 
could be calculated by $m_d \times C(8,K)$, where $m_d =3$ and $K=$ 1, 2 and 3.

As seen from Figure \ref{fig:SALUSIR_ex1_percentage}, the performance of the proposed network SALU"SIR" 
is the same as that of the continuous Hopfield Network for 1-Hamming distance case ($\%100$ for both networks) and 
is slightly higher than that of the Hopfield Network for 2 distance case. 
%However, it's known that the performance of Hopfield network may highly depend on the weight matrices. 
%For example, it's observed that for ${\mathbf A} = - 3{\mathbf I}$, 
%the performance of Hopfield Network is slightly better than the proposed network for the same 
%weights ${\mathbf W}$ and ${\mathbf A}$. 
%So, our test simulation results
%suggest that the proposed network SALU''SIR'' and the Hopfield network, in general, 
%gives comparable performances in many cases. To investigate when either one ourperforms the other one 
%would be an interesting future research item.

\vspace{0.2cm}
\emph{Example 2:}
\vspace{0.2cm}

This example is taken from example 2 in \cite{Uykan08a}. The desired prototype vectors 
as well as the obtained weight matrix ${\bf W}$  are shown in Appendix I.
The other network paramaters are chosen as in example 1: $\sigma = 2$ and $\rho = -2$ .

\begin{figure}[htbp]
  \begin{center}
   \epsfxsize=24.0em    % scale of the figure
\leavevmode\epsffile{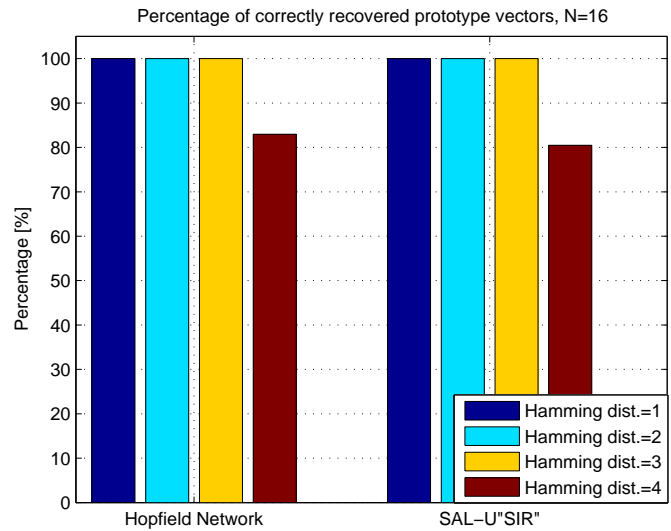}
   \vspace{-1em}        % eliminate extra empty at the bottom of the figure
  \end{center}
 \caption{ The figure 
shows percentage of correctly recovered desired patterns for 
all possible initial conditions in example 2 for the proposed SALU-"SIR"  
as compared to traditional Hopfield network with 16 neurons. }
\label{fig:SALUSIR_ex2_percentage} 
\end{figure}

The Figure \ref{fig:SALUSIR_ex2_percentage} shows the percentages of correctly recovered desired patterns for 
all possible initial conditions $\mathbf{ x }(t=0) \in (-1,+1)^{16}$, in the proposed SALU"SIR" 
as compared to the traditional Hopfield network.

The total number of different possible combinations for the initial conditions for this example is
64, 480 and 2240 and 7280 for 1, 2, 3 and 4-Hamming distance cases respectively, which
could be calculated by $m_d \times C(16,K)$, where $m_d =4$ and $K=$ 1, 2, 3 and 4. 

As seen from Figure \ref{fig:SALUSIR_ex2_percentage} the performance of the proposed network SALU"SIR"
is the same as that of Hopfield Network for 1, 2 and 3-Hamming distance cases ($\%100$ for both networks), 
and gives comparable performance with the Hopfield Network for 4-Hamming distance case.  

\subsection{Discrete-time examples}

In this section, we present two examples, one with 8 neurons (Example 3) and one with 16 neurons (Example 4).
The traditional discrete Hopfield network \cite{Hopfield85}, shown in the following, is used as a reference network:

\begin{equation} \label{eq:discreteHopfield}
{\bf x}^{k+1} = sign \Big( {\bf W} {\bf x}^{k}  \Big)
\end{equation}

where ${\bf W}$ is the weight matrix and ${\bf x}^{k}$ is the state at time $k$, and at most one state is
updated at a step. 

In the simulations in this subsection, we also examine the following version of the DSAL-U''SIR'' 
for comparison reasons: 

\begin{eqnarray}
{\mathbf x}(k+1) & = &( -\rho {\bf I} + {\bf W}  ) {\mathbf x}(k ) 
             g ( | {\bf \theta}^{vec}(t) - {\bf \theta}^{ult, vec} | )  \label{eq:SAL-USIR2a} \\
	     {\mathbf y}(k) & = & sign( {\mathbf x}(k) )   \label{eq:SAL-USIR2b}
\end{eqnarray} 

where ${\bf I}$ is the identity matrix, $ 1 > \rho > 0$ and ${\bf W}$ is defined 
in eq.(\ref{eq:matA_W_b}), 
and ${\mathbf y}(k)$ is the output of the network. It can be shown that the above network is stable
using the steps in DSAL-U''SIR'' in previous section. Here, we omit the proof for the sake of brevity 
and present only the results for comparison reasons. 

Let's denote the original network in eqs.(\ref{eq:SAL-USIR1}) - (\ref{eq:SAL-USIR1n}) as DSAL-U''SIR''1, and 
let's call the network in eq.(\ref{eq:SAL-USIR2a})-(\ref{eq:SAL-USIR2b}) as DSAL-U"SIR"2. 
%and for comparison reasons, we examine the performances of both networks. 

\vspace{0.2cm}
\emph{Example 3:}
\vspace{0.2cm}

The desired prototype vectors are given in the raws of the following matrix 

\begin{equation} \label{eq:ex1_D_ds}
{\mathbf D} =
\left[
\begin{array}{c c c c c c c c}
1   &   1   & 1  &  1  & -1  &  -1  & -1  & -1  \\
1   &   1   & -1  &  -1  & 1  &  1  & -1  & -1  
%1   &   -1  &  1  &  -1  & 1  &  -1  & 1  & -1 
\end{array}
\right]
\end{equation}

The weight matrices $r \bf{ I }$ and $\bf{ W }$, and the threshold vector $\bf{ b }$ 
are obtained as follows 
by using the outer-products-based design mentioned above 
and $\rho$ is chosen as -1 and for the DSALU-U''SIR''2 network, $\rho=0.5$.

\begin{eqnarray} \label{eq:matA_W_b_ex3}
r {\mathbf I} & = & 2{\mathbf I},  \\ 
{\mathbf W}   & = & 
\left[
\begin{array}{c c c c c c c c}
0 & 2 & 0 & 0 & 0 & 0 &-2 &-2 \\
2 & 0 & 0 & 0 & 0 & 0 &-2 &-2 \\
0 & 0 & 0 & 2 &-2 &-2 & 0 & 0 \\
0 & 0 & 2 & 0 &-2 &-2 & 0 & 0 \\
0 & 0 &-2 &-2 & 0 & 2 & 0 & 0 \\
0 & 0 &-2 &-2 & 2 & 0 & 0 & 0 \\
-2 &-2 & 0 & 0 & 0 & 0 & 0 & 2 \\
-2 &-2 & 0 & 0 & 0 & 0 & 2 & 0
\end{array}
\right],  \\
{\mathbf d} & = & {\mathbf 0}
\end{eqnarray}

%where ${\mathbf I}$ shows the identity matrix of dimension $N$ by $N$.

\begin{figure}[htbp]
  \begin{center}
   \epsfxsize=24.0em    % scale of the figure
\leavevmode\epsffile{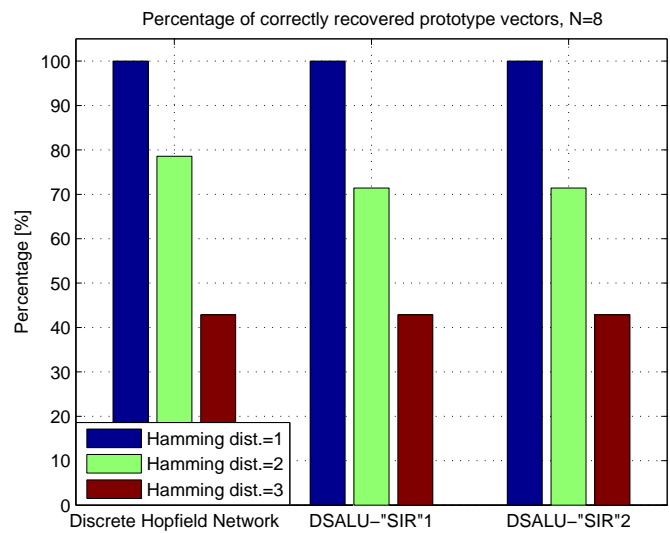}
   \vspace{-1em}        % eliminate extra empty at the bottom of the figure
  \end{center}
 \caption{ The figure
shows percentage of correctly recovered desired patterns for
all possible initial conditions in example 3 for the proposed DSALU-''SIR''1 and DSALU-''SIR''2 
as compared to traditional discrete Hopfield network with 8 neurons. }
\label{fig:DSALUSIR_ex1_percentage}
\end{figure}

The Figure \ref{fig:DSALUSIR_ex1_percentage} shows the percentages of correctly recovered desired patterns for 
all possible initial conditions $\mathbf{ x }(k=0) \in (-1,+1)^N$, in the proposed DSALU-"SIR"1 and DSALU-''SIR''2 
as compared to the traditional discrete Hopfield network in (\ref{eq:discreteHopfield}). 

As seen from Figure \ref{fig:DSALUSIR_ex1_percentage}, the performances of the DSALU-''SIR''1 and 2
are the same as that of the discrete-time Hopfield Network for 1-Hamming distance case ($\%100$ for both networks) and are comparable for 2 and 3-Hamming distance cases respectively.

\vspace{0.2cm}
\emph{Example 4:}
\vspace{0.2cm}

The desired prototype vectors as well as the obtained weight matrices 
are given in in Appendix II (eq.(\ref{eq:ex4_D})). 

For matrix $r{\bf I}$, $\rho$ is chosen as -2.
The other network paramaters 
are chosen as in example 3.

\begin{figure}[htbp]
  \begin{center}
   \epsfxsize=24.0em    % scale of the figure
\leavevmode\epsffile{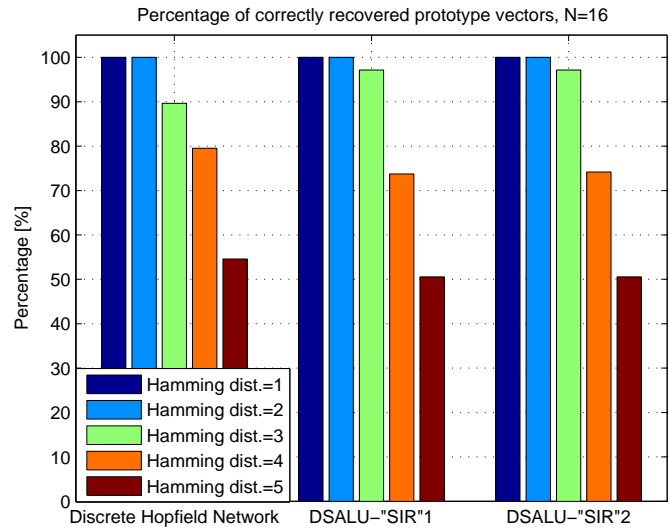}
   \vspace{-1em}        % eliminate extra empty at the bottom of the figure
  \end{center}
 \caption{ The figure
shows percentage of correctly recovered desired patterns for
all possible initial conditions in example 4 for the proposed 
DSALU-''SIR''1 and DSALU-''SIR''2 
as compared to the traditional discrete Hopfield network with 16 neurons. }
\label{fig:DSALUSIR_ex2_percentage}
\end{figure}

The Figure \ref{fig:DSALUSIR_ex2_percentage} shows the percentages of correctly recovered desired patterns for 
all possible initial conditions $\mathbf{ x }(k=0) \in (-1,+1)^{16}$, in the proposed DSALU"SIR"1 and 2 
as compared to the traditional Hopfield network.  

The total number of different possible combinations for the initial conditions for this example is
64, 480 and 2240 and 7280 for 1, 2, 3 and 4-Hamming distance cases respectively. 
%which could be calculated by $m_d \times C(16,K)$, where $m_d =4$ and $K=$ 1, 2, 3 and 4. 

As seen from Figure \ref{fig:DSALUSIR_ex2_percentage} the performance of the proposed networks DSALU-"SIR"1 and 2  
are the same as that of Hopfield Network for 1 and 2-Hamming distance cases ($\%100$ for both networks), and 
are comparable for 3,4 and 5-Hamming distance cases respectively.

\section{Conclusions  \label{Section:CONCLUSIONS}}

In this paper, we present and analyse two Hopfield-like nonlinear networks, 
%called SALU-SIR and DSALU-SIR 
in continuous-time and discrete-time respectively. 
The proposed network is based on an autonomous linear system with a symmetric weight matrix, 
which is designed to be unstable, 
and a nonlinear function stabilizing the whole network 
thanks to a manipulated state variable. 
This variable is observed to be equal to the traditional 
Signal-to-Interference Ratio (SIR) definition in telecommunications engineering.

The underlying linear system of the proposed continuous-time network is 
$\dot{ {\mathbf x}} = {\mathbf B} {\mathbf x}$ where 
{\bf B} is a real symmetric matrix whose 
diagonal elements are fixed to a constant. 
The nonlinear function, on the other hand, is based on the defined system variables 
called ``SIR''s. We also show that 
the ``SIR''s of all the states converge to a constant value, called ``system-specific Ultimate SIR''; 
which is equal to $\frac{r}{\lambda_{max}}$ where $r$ is the diagonal element
of matrix ${\bf B}$ 
and $\lambda_{max}$ is the maximum (positive) eigenvalue of diagonally-zero
matrix $({\bf B} - r{\bf I})$, where ${\bf I}$ denotes the identity matrix. 
The same result is obtained in its discrete-time version as well. 

Computer simulations for binary associative memory design problem show the effectiveness of the
proposed network as compared to the traditional Hopfield Networks.

%\smallskip
\vspace{0.2cm}

%\newpage
%
\section*{Acknowledgments}

This work was supported in part by Academy of Finland and Research Foundation (Tukis\"{a}\"{a}ti\"{o}) of Helsinki 
University of Technology, Finland. 

The author would also like to thank the anonymous four reviewers for 
their valuable comments which helped in improving the 
structure and the content of the paper. 

\nocite{*}
\bibliographystyle{IEEE}
%%%%%\bibliography{bib-file}  % commented if *.bbl file included, as
%%%%%see below

%%%%%%%%%%%%%%%%% BIBLIOGRAPHY IN THE LaTeX file !!!!! %%%%%%%%%%%%%%%%%%%%%%%%
%% This is nothing else than the IEEEsample.bbl file that you would
%%
%% obtain with BibTeX: you do not need to send around the *.bbl file
%%
%%---------------------------------------------------------------------------%%
%
%\begin{thebibliography}{1}
%\bibitem{LaTeX}
%Leslie Lamport,
%\newblock {\em A Document Preparation System: \LaTeX, User's Guide and
%  Reference Manual},
%\newblock Addison Wesley Publishing Company, 1986.
%\end{thebibliography}

\vspace{3cm}

%\bibliographystyle{plain}
%\bibliography{doctorate_short}

\newpage
\listoffigures

\newpage

\section*{Appendix I}

In Example 2, the matrix which has the desired prototype vectors as its raws is 

\begin{equation} \label{eq:ex2_D}
{\mathbf D} =
\left[
\begin{array}{c c c c c c c c c c c c c c c c}
1 & 1 & 1 & 1 & 1 & 1 & 1 & 1 & -1 & -1 & -1 & -1 & -1 & -1 & -1 & -1 \\
1 & 1 & 1 & 1 & -1 & -1 & -1 & -1 & 1 & 1 & 1 & 1 & -1 & -1 & -1 & -1 \\
1 & 1 & -1 & -1 & 1 & 1 & -1 & -1 & 1 & 1 & -1 & -1 & 1 & 1 & -1 & -1 \\
1 & -1 & 1 & -1 & 1 & -1 & 1 & -1 & 1 & -1 & 1 & -1 & 1 & -1 & 1 & -1
\end{array}
\right]
\end{equation}

\vspace{0.4cm}

In Example 2, the weight matrices $r{\bf I}$ and ${\bf W}$, which are obtained by the outer products based design as explained 
in Section \ref{Subsection:HebbBasedLearning}, are as follows:

\newpage

\begin{eqnarray} \label{eq:matA_W_b_ex2}
r {\mathbf I} & = & 2 {\mathbf I} \\
{\mathbf W} & = &
\left[
\begin{array}{c c c c c c c c c c c c c c c c}
 0  &  2  &  2  &  0  &  2  &  0  &  0  & -2  &  2  &  0  &  0  & -2  &  0  & -2  & -2  & -4 \\
 2  &  0  &  0  &  2  &  0  &  2  & -2  &  0  &  0  &  2  & -2  &  0  & -2  &  0  & -4  & -2 \\
 2  &  0  &  0  &  2  &  0  & -2  &  2  &  0  &  0  & -2  &  2  &  0  & -2  & -4  &  0  & -2 \\
 0  &  2  &  2  &  0  & -2  &  0  &  0  &  2  & -2  &  0  &  0  &  2  & -4  & -2  & -2  &  0 \\
 2  &  0  &  0  & -2  &  0  &  2  &  2  &  0  &  0  & -2  & -2  & -4  &  2  &  0  &  0  & -2 \\
 0  &  2  & -2  &  0  &  2  &  0  &  0  &  2  & -2  &  0  & -4  & -2  &  0  &  2  & -2  &  0 \\
 0  & -2  &  2  &  0  &  2  &  0  &  0  &  2  & -2  & -4  &  0  & -2  &  0  & -2  &  2  &  0 \\
-2  &  0  &  0  &  2  &  0  &  2  &  2  &  0  & -4  & -2  & -2  &  0  & -2  &  0  &  0  &  2 \\
 2  &  0  &  0  & -2  &  0  & -2  & -2  & -4  &  0  &  2  &  2  &  0  &  2  &  0  &  0  & -2 \\
 0  &  2  & -2  &  0  & -2  &  0  & -4  & -2  &  2  &  0  &  0  &  2  &  0  &  2  & -2  &  0 \\
 0  & -2  &  2  &  0  & -2  & -4  &  0  & -2  &  2  &  0  &  0  &  2  &  0  & -2  &  2  &  0 \\
-2  &  0  &  0  &  2  & -4  & -2  & -2  &  0  &  0  &  2  &  2  &  0  & -2  &  0  &  0  &  2 \\
 0  & -2  & -2  & -4  &  2  &  0  &  0  & -2  &  2  &  0  &  0  & -2  &  0  &  2  &  2  &  0 \\
-2  &  0  & -4  & -2  &  0  &  2  & -2  &  0  &  0  &  2  & -2  &  0  &  2  &  0  &  0  &  2 \\
-2  & -4  &  0  & -2  &  0  & -2  &  2  &  0  &  0  & -2  &  2  &  0  &  2  &  0  &  0  &  2 \\
-4  & -2  & -2  &  0  & -2  &  0  &  0  &  2  & -2  &  0  &  0  &  2  &  0  &  2  &  2  &  0 
\end{array}
\right] \nonumber \\
 &  &  
\end{eqnarray}

%\newpage

\section*{Appendix II}

In Example 4, the matrix which has the desired prototype vectors as its raws is 

\begin{equation} \label{eq:ex4_D}
{\mathbf D} =
\left[
\begin{array}{c c c c c c c c c c c c c c c c}
1 & 1 & 1 & 1 & 1 & 1 & 1 & 1 & -1 & -1 & -1 & -1 & -1 & -1 & -1 & -1 \\
1 & 1 & 1 & 1 & -1 & -1 & -1 & -1 & 1 & 1 & 1 & 1 & -1 & -1 & -1 & -1 \\
1 & 1 & -1 & -1 & 1 & 1 & -1 & -1 & 1 & 1 & -1 & -1 & 1 & 1 & -1 & -1
%1 & -1 & 1 & -1 & 1 & -1 & 1 & -1 & 1 & -1 & 1 & -1 & 1 & -1 & 1 & -1
\end{array}
\right]
\end{equation}

In Example 4, the weight matrices $r{\bf I}$ and ${\bf W}$ obtained are as follows:

\begin{eqnarray} \label{eq:matA_W_b_ex4}
r {\mathbf I} & = & 3 {\mathbf I}, \nonumber \\
{\mathbf W} & = &
\left[
\begin{array}{c c c c c c c c c c c c c c c c}
0 &   3 &   1 &   1 &   1 &   1 &  -1 &  -1 &   1 &   1 &  -1 &  -1 &  -1 &  -1 &  -3 &  -3 \\
3 &   0 &   1 &   1 &   1 &   1 &  -1 &  -1 &   1 &   1 &  -1 &  -1 &  -1 &  -1 &  -3 &  -3 \\
1 &   1 &   0 &   3 &  -1 &  -1 &   1 &   1 &  -1 &  -1 &   1 &   1 &  -3 &  -3 &  -1 &  -1 \\
1 &   1 &   3 &   0 &  -1 &  -1 &   1 &   1 &  -1 &  -1 &   1 &   1 &  -3 &  -3 &  -1 &  -1 \\
1 &   1 &  -1 &  -1 &   0 &   3 &   1 &   1 &  -1 &  -1 &  -3 &  -3 &   1 &   1 &  -1 &  -1 \\
1 &   1 &  -1 &  -1 &   3 &   0 &   1 &   1 &  -1 &  -1 &  -3 &  -3 &   1 &   1 &  -1 &  -1 \\
-1 &  -1 &   1 &   1 &   1 &   1 &   0 &   3 &  -3 &  -3 &  -1 &  -1 &  -1 &  -1 &   1 &   1 \\
-1 &  -1 &   1 &   1 &   1 &   1 &   3 &   0 &  -3 &  -3 &  -1 &  -1 &  -1 &  -1 &   1 &   1 \\
1 &   1 &  -1 &  -1 &  -1 &  -1 &  -3 &  -3 &   0 &   3 &   1 &   1 &   1 &   1 &  -1 &  -1 \\
1 &   1 &  -1 &  -1 &  -1 &  -1 &  -3 &  -3 &   3 &   0 &   1 &   1 &   1 &   1 &  -1 &  -1 \\
-1 &  -1 &   1 &   1 &  -3 &  -3 &  -1 &  -1 &   1 &   1 &   0 &   3 &  -1 &  -1 &   1 &   1 \\
-1 &  -1 &   1 &   1 &  -3 &  -3 &  -1 &  -1 &   1 &   1 &   3 &   0 &  -1 &  -1 &   1 &   1 \\
-1 &  -1 &  -3 &  -3 &   1 &   1 &  -1 &  -1 &   1 &   1 &  -1 &  -1 &   0 &   3 &   1 &   1 \\
-1 &  -1 &  -3 &  -3 &   1 &   1 &  -1 &  -1 &   1 &   1 &  -1 &  -1 &   3 &   0 &   1 &   1 \\
-3 &  -3 &  -1 &  -1 &  -1 &  -1 &   1 &   1 &  -1 &  -1 &   1 &   1 &   1 &   1 &   0 &   3 \\
-3 &  -3 &  -1 &  -1 &  -1 &  -1 &   1 &   1 &  -1 &  -1 &   1 &   1 &   1 &   1 &   3 &   0 
\end{array}
\right],   \nonumber 
\end{eqnarray}

%\listoffigures

\end{document}